\documentclass[a4paper,fleqn,usenatbib]{mnras}
\usepackage{newtxtext,newtxmath}
\usepackage[T1]{fontenc}
\usepackage{ae,aecompl}
\usepackage{graphicx}	
\usepackage{amsmath}	

\usepackage{amssymb}	
\usepackage{mathrsfs}
\usepackage{longtable}
\usepackage{hyperref}
\usepackage[font=small]{caption}
\usepackage{bm}

\title[SN environments]{Toward a better understanding of supernova environments: a study of SNe~2004dg and 2012P in NGC~5806 with HST and MUSE}

\author[Sun et al.]{Ning-Chen Sun$^1$\thanks{E-mail: n.sun@sheffield.ac.uk}, 
Justyn R. Maund$^1$,
Paul A. Crowther$^1$,
Xuan Fang$^2$
\newauthor
and Emmanouil Zapartas$^3$ \\
1 Department of Physics and Astronomy, University of Sheffield, Hicks Building, Hounsfield Road, Sheffield S3 7RH, UK \\
2 Key Laboratory of Optical Astronomy, National Astronomical Observatories of the Chinese Academy of Sciences (NAOC), Beijing 100101, China  \\
3 Geneva Observatory, University of Geneva, Chemin des Maillettes 51, 1290 Versoix, Switzerland}

\date{Accepted XXX. Received YYY; in original form ZZZ}

\pubyear{2020}

\begin{document}
\label{firstpage}
\pagerange{\pageref{firstpage}--\pageref{lastpage}}
\maketitle

\defcitealias{F16}{F16}
\newcommand*{\iipmass}{\ensuremath{10.0^{+0.3}_{-0.2}}}
\newcommand*{\iibmass}{\ensuremath{15.2^{+2.0}_{-1.0}}}

\begin{abstract}

Core-collapse supernovae (SNe) are the inevitable fate of most massive stars. Since most stars form in groups, SN progenitors can be constrained with information of their environments. It remains challenging to accurately analyse the various components in the environment and to correctly identify their relationships with the SN progenitors. Using a combined dataset of VLT/MUSE spatially-resolved integral-field-unit (IFU) spectroscopy and HST/ACS+WFC3 high-spatial resolution imaging, we present a detailed investigation of the environment of the Type~II-P SN~2004dg and Type~IIb SN~2012P. The two SNe occurred in a spiral arm of NGC~5806, where a star-forming complex is apparent with a giant H~\textsc{ii} region. By modelling the ionised gas, a compact star cluster and the resolved stars, we derive the ages and extinctions of stellar populations in the vicinity of the SNe. The various components are consistent with a sequence of triggered star formation as the spiral density wave swept through their positions. For SNe~2004dg and 2012P, we identify their host stellar populations and derive initial masses of \iipmass~$M_\odot$ and  \iibmass~$M_\odot$ for their progenitors, respectively. Both results are consistent with those from pre-explosion images or nebular-phase spectroscopy. SN~2012P is spatially coincident but less likely to be coeval with the star-forming complex. As in this case, star formation bursts on small scales may appear correlated if they are controlled by any physical processes on larger scales; this may lead to a high probability of chance alignment between older SN progenitors and younger stellar populations.
\end{abstract}


\section{Introduction}
\label{intro.sec}

%
Core-collapse SNe are spectacular explosions of massive stars at the end of their lives. Understanding the progenitors of different SN subtypes is an important topic in the study of SN explosion and massive stellar evolution. Direct progenitor detections have been successful for only a limited number of SNe \citep[e.g.][]{Smartt2004, Maund2005, Maund2009, Smartt2009, Eldridge2013, Davies2018, VanDyk2018}. Relatively reliable estimates for the progenitors' properties can also be obtained by modelling the SN spectra at nebular phase, when the SN ejecta becomes optically thin and the material from the core can be observed \citep{Jerkstrand2015}. However, SNe become very faint at nebular phase so it is challenging to acquire high-quality spectra. In the era of transient astronomy, there is an urgent need for alternative techniques to identify their progenitors as we are witnessing a rapid increase in the numbers of newly discovered SNe and SN subtypes.

Analysing the environments of SNe can provide important information for their progenitors. Most stars are formed in groups, and stars in each group have very similar ages and uniform metallicities \citep[e.g.][]{L03.ref}. Massive stars can only travel a limited distance during their short lifetimes; as a result, most of the observed SNe should still be very close to the host stellar populations. Thus, the properties of SN progenitors can be inferred by analysing their environments. In particular, their lifetimes should correspond to the ages of their host stellar populations.

Note that, while environment analysis derives the lifetime, the pre-explosion images or nebular-phase spectroscopy reflect the final state of the SN progenitors. In case of any binary interaction, the relation between the final state and the stellar lifetime may be different from that expected from single stellar evolution \citep{Z19, Z20}. Thus, environment analysis, combined with the direct methods, provides important insights into the full evolution history for the progenitors.

The analysis of SN environments has been conducted with various tracers and based on different types of observations. Recently, IFU spectroscopy has been used as a powerful tool in SN environment studies \citep[e.g.][]{K13a, K13b, K18, G14, G16a, G16b, G18, X18, X19, Schady2019, Lyman2020}. The nebular emission lines are very sensitive to and can be used to derive the ages of the ionising stellar populations since they are powered by young massive stars' ionising radiation at ultraviolet (UV) wavelengths. Properties of the SN progenitors can then be accurately inferred if they are coeval with the ionising stellar populations. 

However, deriving the ages is not trivial since the spectrum of an H~\textsc{ii} region may also depend on many other parameters, such as the covering factor (the fraction of sky covered by the gas as viewed from the ionising sources), optical depth, abundances of individual elements, and dust within the gas. All these parameters remain highly uncertain but are very important in the modelling of the ionised gas \citep[e.g.][]{X18, X19, Schady2019}. Moreover, the assumption of co-evolution between the SN progenitors and the ionising sources may not be valid in some circumstances. Limited by the spatial resolution, it is hard to tell whether the SNe may actually exhibit an offset from the ionised gas and/or the ionising sources \citep[e.g.][]{Maund2016}; it is also possible that the SN progenitors may come from significantly older stellar populations in chance alignment.

High-spatial resolution imaging has also been extensively used to study SN environments. With the HST, one can resolve the individual stars or star clusters in galaxies out to several tens Mpc. A handful of SNe are found to be spatially coincident with and very likely to be hosted by compact star clusters. In this case, their progenitors' properties can be reliably derived by analysing the host clusters' spectral energy distributions (SEDs; e.g. \citealt{MA2004}; \citealt{S20b}). Somewhat surprisingly, most SNe do not occur in compact star clusters. Their host clusters may have already dissolved into the field or their progenitors are born in much looser agglomerations of stars. Thus, one has to investigate the resolved stellar populations in their environments in order to derive the properties of their progenitors \citep[e.g.][]{Gogarten2009, Murphy2011, Shivvers2017, Diaz2018, Auchettl2019, S20a}.

Compared with star clusters, the analysis of resolved stellar populations is much more complicated since the observed stars may have different ages and/or extinctions. They are usually modelled by maximising the likelihood of a linear combination of model colour-magnitude diagrams (CMDs) to the observed CMD \citep[e.g.][]{Murphy2011} or by directly comparing the observed magnitudes with model predictions for individual stars \citep[e.g.][]{Maund2016}. Yet, the age estimate for resolved stars may depend on the uncertain metallicity and/or the treatment of extinction (e.g. whether all stars have the same extinction) due to the age-metallicity-extinction degeneracy. Moreover, the resolved stars surrounding a SN often have multiple age components \citep{Murphy2011, Maund2017, Maund2018}; without additional information (e.g. kinematics, spatial distribution along the line of sight), it may be difficult to identify their mutual relationships and the host stellar population that is coeval with the SN progenitor.

Thus, for a given SN, we need to overcome two main challenges toward a comprehensive understanding of its environment and a reliable inference for its progenitor: (1) to reveal the various components in the environment and accurately derive their physical parameters; and (2) to correctly identify their relationships with each other and, in particular, with the SN progenitor. We suggest that this can be achieved by combining spatially-resolved IFU spectroscopy and high-spatial resolution imaging. With a simultaneous analysis of both the gaseous and the stellar components, one can resolve degeneracies that plague the individual techniques and address the complexities of the environments themselves.

In this paper, we aim to (1) present a general framework of environment analysis with the combined datasets, and to (2) check whether the analysis can reliably constrain the SN progenitors by comparing with results from direct methods. Here we look at the cases of SNe 2004dg and 2012P, which occurred in the nearby spiral galaxy NGC~5806. SN~2004dg is a hydrogen-rich Type~II-P SN arising from the explosion of a red supergiant (RSG). Its progenitor was not detected on pre-explosion images; \citet{Smartt2009} derived a corresponding initial mass limit of $M_{\rm ini}$~$\leq$~12~$M_\odot$, which was then updated to $M_{\rm ini}$~$\leq$~9.5$^{+0.7}$~$M_\odot$ with a new bolometric correction \citep{Davies2018}. SN~2012P, also named PTF12os, has been carefully studied by \citet[][\citetalias{F16} hereafter]{F16} for its light curve and spectral evolution. It is a Type~IIb SN since the hydrogen features quickly disappeared within $\sim$25~days after explosion. With nebular-phase spectroscopy, \citetalias{F16} obtained an oxygen mass of 0.73~$M_\odot$ that corresponds to a progenitor of $M_{\rm ini}$~=~15~$M_\odot$. The SN sites have been imaged by the Hubble Space Telescope (HST) and their IFU spectroscopy has been obtained by Multi-Unit Spectroscopic Explorer (MUSE) mounted on the Very Large Telescope (VLT). Since the two SNe are so close to each other, their analysis can be done homogeneously with the same dataset and with a single distance modulus ($\mu$~=~32.14~$\pm$~0.20~mag; i.e. 26.8$^{+2.6}_{-2.4}$~Mpc; \citealt{Tully2013}; consistent with that used in \citetalias{F16}).

\section{Data}
\label{data.sec}

\begin{table}
\center
\caption{HST Observations of SN~2004dg and SN~2012P}
\begin{tabular}{ccccc}
\hline
\hline
Program & Date & Instrument & Filter & Exposure \\
ID & (UT) &   &    & Time (s) \\
\hline
9788$^{\rm a}$ 
& 2004-04-03 &  ACS/WFC & F658N &  700.0 \\
& 2004-04-03 &  ACS/WFC & F814W &  120.0 \\
\hline
10187$^{\rm b}$ 
& 2005-03-10 &  ACS/WFC & F435W &  1600.0 \\
& 2005-03-10 &  ACS/WFC & F555W &  1400.0 \\
& 2005-03-10 &  ACS/WFC & F814W &  1700.0 \\
\hline
13822$^{\rm c}$ 
& 2015-06-30 &  WFC3/UVIS & F225W &  8865.0 \\
& 2015-06-30 &  ACS/WFC & F814W &  2345.0 \\
\hline
15152$^{\rm d}$ 
& 2019-06-25 &  WFC3/UVIS & F555W &  5437.0 \\
& 2019-06-25 &  WFC3/UVIS & F438W &  5408.0 \\
& 2019-06-25 &  ACS/WFC & F814W &  2325.0 \\
\hline
\multicolumn{5}{l}{PIs: (a) L. Ho; (b) S. Smartt; (c) G. Folatelli; (d) S. Van Dyk.}\\
\end{tabular}
\label{obs.tab}
\end{table}

Spatially-resolved IFU spectra of NGC~5806 were obtained in December 2016 with VLT/MUSE as part of the program ``The MUSE Atlas of Disks (097.B-0165; PI: C.~M.~Carollo)". The observations were conducted in the wide-field mode (WFM) with extended wavelength setting, covering a spatial extent of 1~$\times$~1~arcmin$^2$ with 0.2~arcsec sampling and a spectral range 4750--9350~{\AA}  with 1.25~{\AA} sampling. The spectral resolution of the instrument ($R$~=~$\lambda$/$\Delta \lambda$) ranges from 1770 at the blue end to 3590 at the red end of the spectrum. The seeing was around 0.7~arcsec during the observations. The observation consists of four on-source exposures for a total of 3600~s. Three 120-s sky frames were also acquired for sky subtraction. The raw data was well processed by the standard pipeline and we retrieved the calibrated data from the ESO data archive (\url{http://archive.eso.org}). At each spaxel, we use a boxcar average of 3~$\times$~3 spaxels centred on that spaxel to increase the signal-to-noise ratio (SNR).

A series of archival observations from the Hubble Space Telescope (HST) are also used in this work, a complete list of which is provided in Table~\ref{obs.tab}. They were conducted by the Wide Field Camera 3 (WFC3) Ultraviolet-Visible (UVIS) channel and the Advanced Camera for Surveys (ACS) Wide Field Channel (WFC). The observations span a long period from April 2004 to June 2019. They also cover a long wavelength range from UV (F225W) to near-infrared (F814W). The images were retrieved from the Mikulski Archive for Space Telescopes (\url{https://archive.stsci.edu/index.html}) and re-calibrated with the \textsc{drizzlepac} package for better image alignment and cosmic-ray removal. The SN sites were also observed by an HST program (9042) with the Wide Field and Planetary Camera 2 (WFPC2) in the F450W and F814W bands. This observation is not used in this work due to its lower spatial resolution. Point source photometry is conducted with the \textsc{dolphot} package \citep{dolphot.ref}.

Relative astrometry between the IFU datacube and the HST images is conducted with 7 common stars in the field. The accuracy corresponds to $\sim$1~ACS/WFC pixel or $\sim$0.25 (raw) MUSE spaxel.

\section{An overview of the SN environment}
\label{overview.sec}

\begin{figure*}
\centering
\includegraphics[width=0.9\linewidth, angle=0]{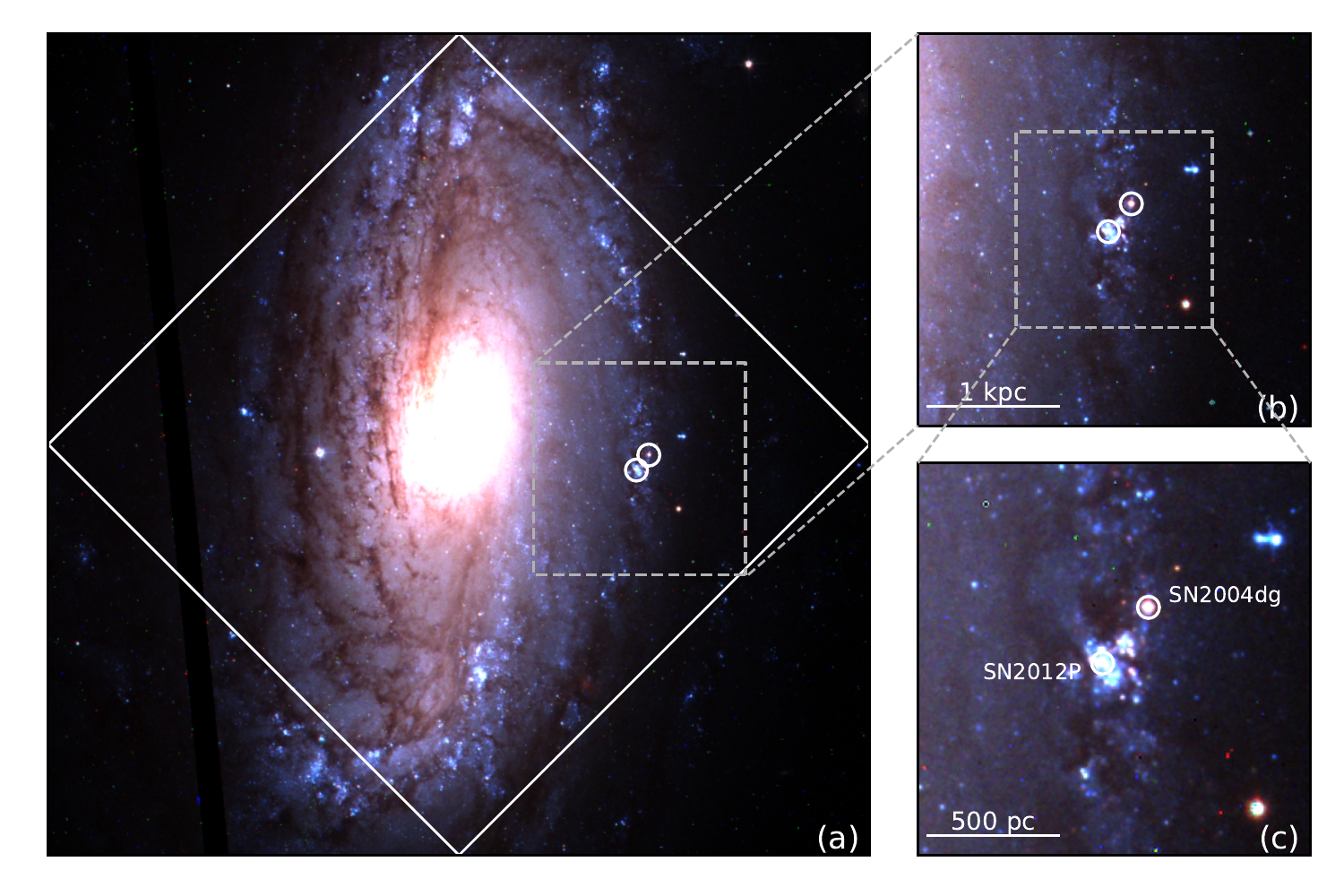}
\caption{Three-colour composite of F435W, F555W and F814W images for NGC~5806 taken by HST/ACS in 2005 (when SN~2004dg was still bright). The white diamond in Panel (a) corresponds to the field of MUSE observations. In all panels, the two circles show the positions of SN~2004dg and SN~2012P. North is up and east is to the left.}
\label{im1.fig}
\includegraphics[width=0.9\linewidth, angle=0]{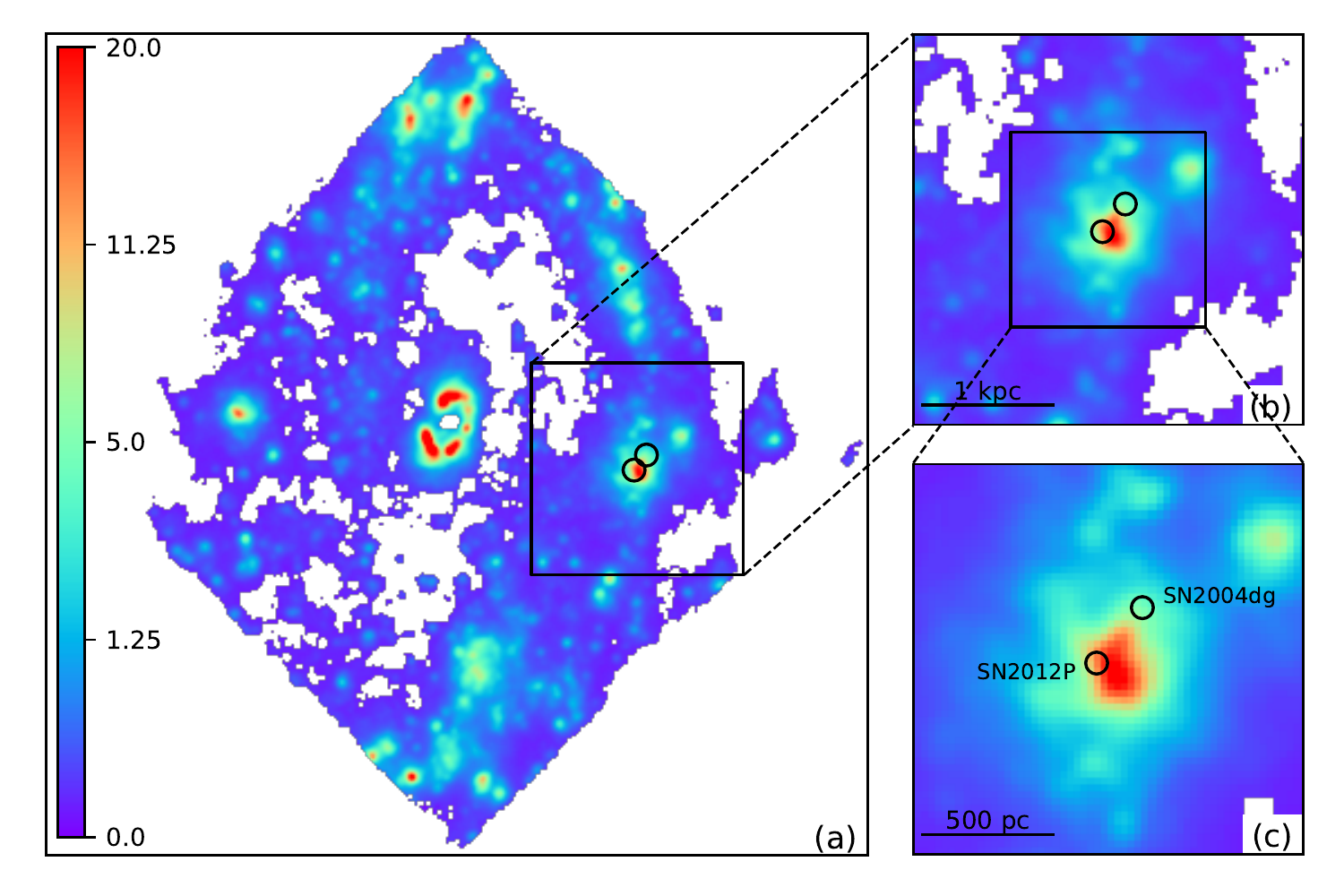}
\caption{Map of H$\alpha$ integrated flux for NGC~5806. The colour bar is in unit of 10$^{-17}$~erg~s$^{-1}$~cm$^{-2}$. The two circles show the positions of SN~2004dg and SN~2012P. North is up and east is to the left.}
\label{im2.fig}
\end{figure*}

\begin{figure*}
\centering
\includegraphics[width=0.9\linewidth, angle=0]{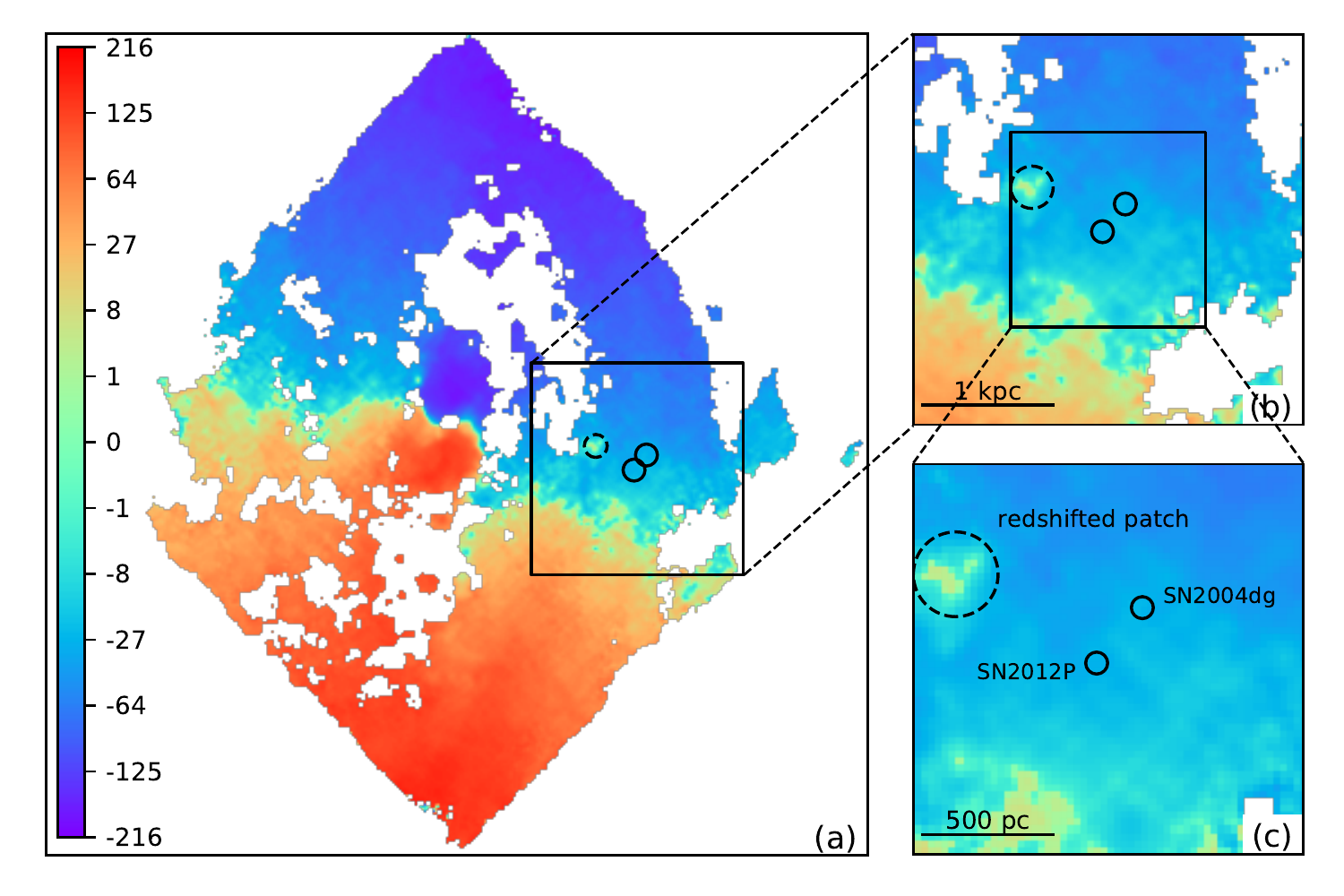}
\caption{Map of H$\alpha$ line central velocity for NGC~5806 relative to the galaxy's systematic recession velocity (1375~km~s$^{-1}$). The colour bar is in unit of km~s$^{-1}$. The dashed circle shows the position of the redshifted patch (see text) and the two solid circles show the positions of SN~2004dg and SN~2012P. North is up and east is to the left.}
\label{im3.fig}
\includegraphics[width=0.9\linewidth, angle=0]{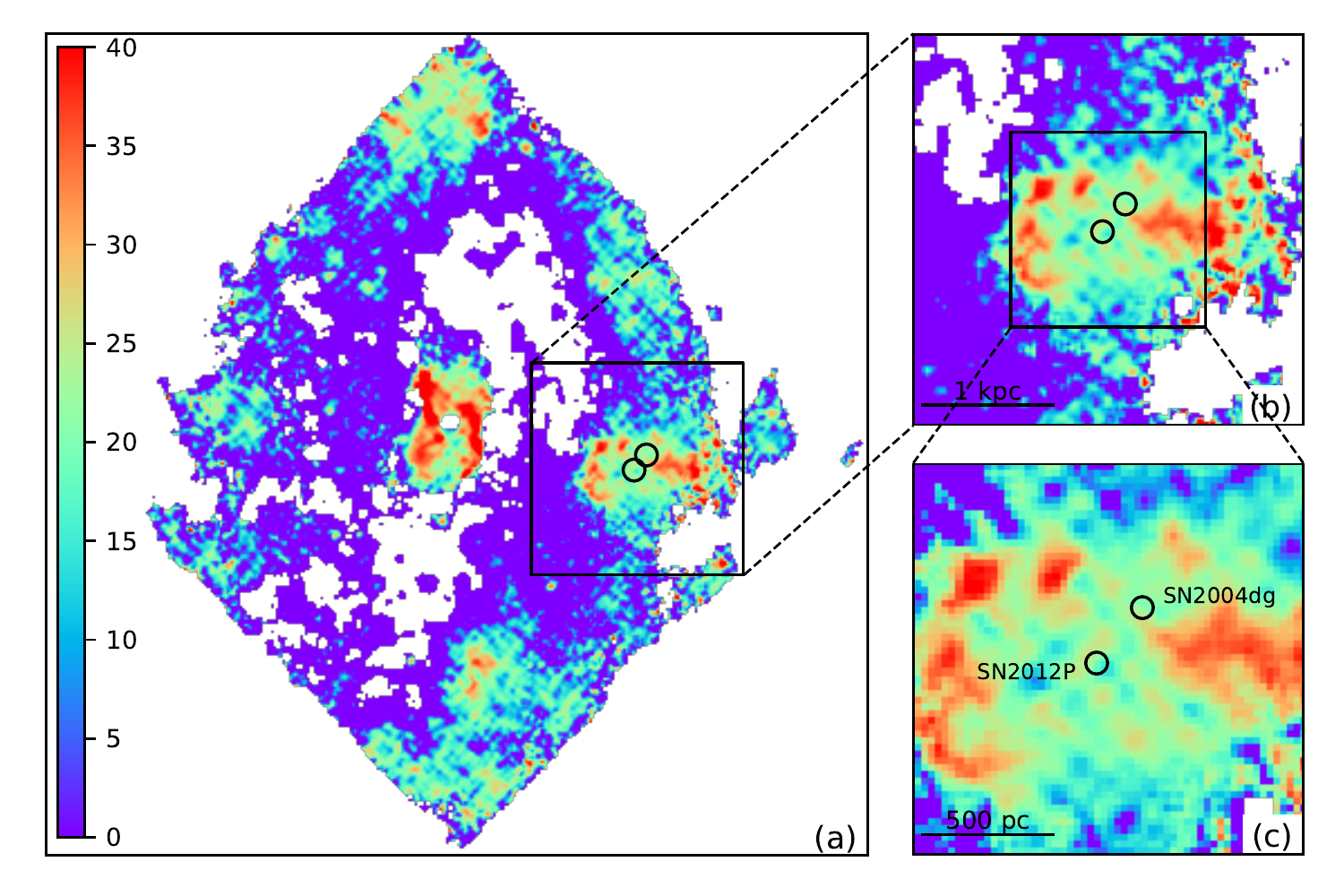}
\caption{Map of H$\alpha$ line width for NGC~5806. An instrumental broadening of 49.3~km~s$^{-1}$ has been removed. The colour bar is in unit of km~s$^{-1}$. The two circles show the positions of SN~2004dg and SN~2012P. North is up and east is to the left.}
\label{im4.fig}
\end{figure*}

\begin{figure*}
\centering
\includegraphics[width=0.9\linewidth, angle=0]{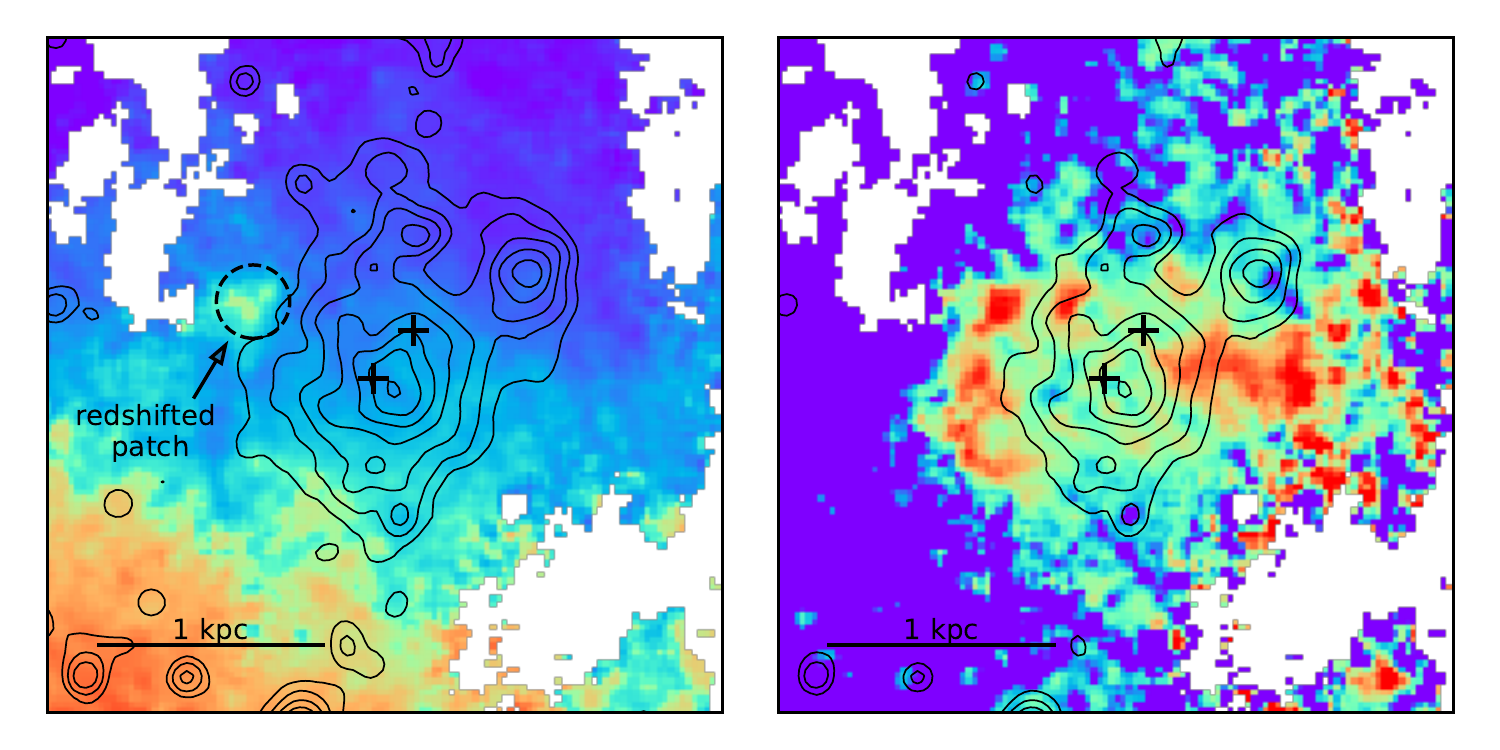}
\caption{Contours of H$\alpha$ integrated flux overlaid on H$\alpha$ central velocity (left) and line width (right) maps (which are the same as Figs.~\ref{im3.fig}b and \ref{im4.fig}b, respectively). The innermost contour corresponds to a flux of 2~$\times$~10$^{-16}$~erg~s$^{-1}$~cm$^{-2}$; the flux decreases by a factor of 2 between every two neighbouring contours from inside to outside. In both panels, the two crosses show the positions of SN~2004dg and SN~2012P. North is up and east is to the left.}
\label{turb.fig}
\end{figure*}

\begin{figure*}
\centering
\includegraphics[width=0.95\linewidth, angle=0]{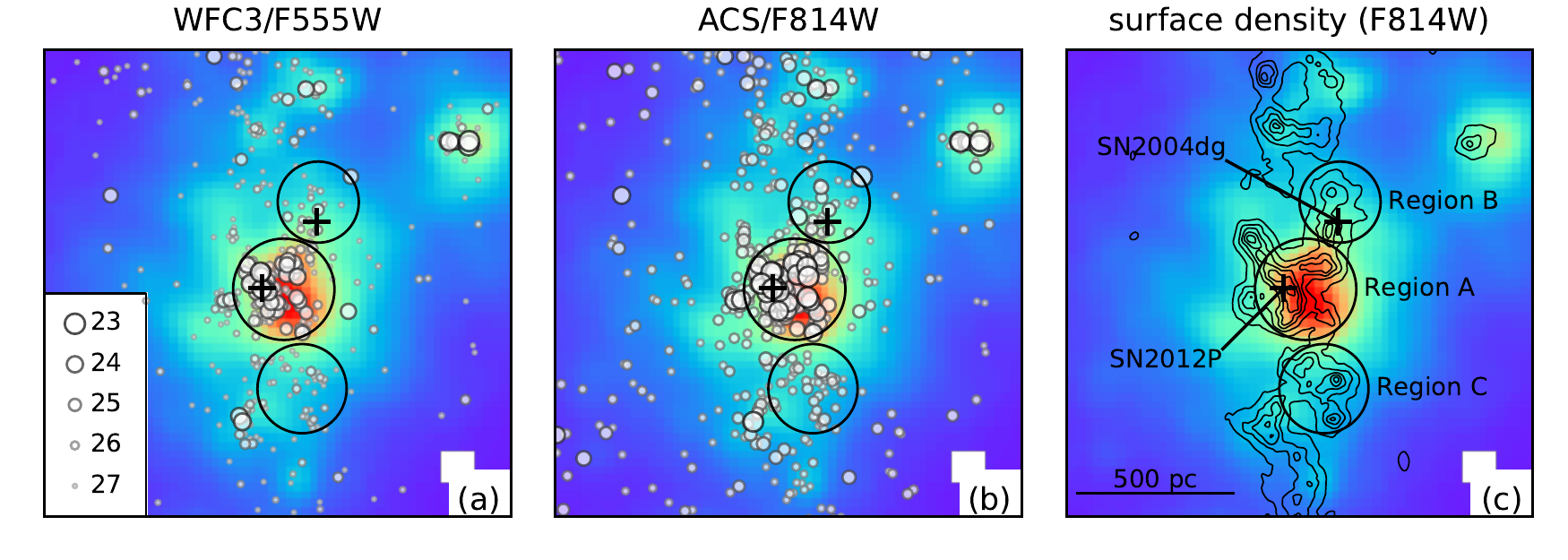}
\caption{(a, b): point sources (stars or star clusters) in the WFC3/F555W and ACS/F814W bands; the symbol size reflects the source's magnitude according to the legend in the first panel; sources with SNR~$<$~10 are not shown to reduce symbol crowding. (c) surface density (estimated with the 10th-nearest neighbour method) contours of F814W sources with SNR~$>$~5; the surface density has a median value of $\Sigma_0$~=~5.11~arcsec$^{-2}$ and a standard deviation of $\sigma$~=~2.74~arcsec$^{-2}$; the contours increase from $\Sigma_0 + 3\sigma$ to $\Sigma_0 + 18\sigma$ from outside to inside in steps of 3$\sigma$. In all panels, the background colour scale shows the H$\alpha$ integrated flux (same as in Fig.~\ref{im2.fig}c); the two crosses show the positions of SN~2004dg and SN~2012P; the three circles define \textit{Regions~A-C}, with which we study the local SN environments (\textit{A} and \textit{B}) or the extinction along the spiral arm (\textit{C}). North is up and east is to the left.}
\label{struct.fig}
\end{figure*}

\subsection{Stellar and gaseous components}
\label{components.sec}

Figure~\ref{im1.fig} shows the HST images of NGC~5806, the host galaxy of SNe~2004dg/2012P. Two spiral arms are clearly revealed by the spatial distribution of young stars; in particular, SNe~2004dg/2012P occurred in one of the spiral arms, where a star-forming complex is apparent with a size of $\sim$300~pc. The complex hosts a large number of bright and blue sources, which could be individual stars or compact star clusters. SN~2004dg shows an offset of $\sim$200~pc from the complex while the projected position of SN~2012P is located inside the complex. \citet{VanDyk2012} and \citetalias{F16} showed that SN~2012P is spatially coincident with a point source; \citetalias{F16} further found this point source to have consistent brightness between pre-explosion and late-time observations. Thus, the point source could be a compact star cluster that hosts SN~2012P's progenitor; alternatively, it may be an unrelated source in chance alignment with the SN position.

The MUSE IFU spectroscopy allows us to further analyse the ionised gas in NGC~5806. A map of H$\alpha$ integrated flux is shown in Fig.~\ref{im2.fig}, derived by fitting Gaussian functions to the emission lines (for simplicity, we do not apply any correction for stellar absorption; note that the emission lines near the SN positions are very strong and any possible stellar absorption can be neglected; see Section~\ref{spec.sec}). The H$\alpha$ emission is prominent within a nuclear starburst ring and along the two spiral arms. At the site of SN~2004dg/SN~2012P, the star-forming complex has photoionised a giant H~\textsc{ii} region, which has a bright core spatially coincident with the complex and an extended envelope with a diameter of $\sim$1~kpc. Among the brightest in NGC5806, this H~\textsc{ii} region has a very high luminosity of $L$(H$\alpha$)~=~4.4~$\times$~10$^{39}$~erg~s$^{-1}$ (corrected for extinction) within the central 1.25~arcsec (i.e. 162~pc; see Section~\ref{spec.sec}); this value lies at the bright end of H~\textsc{ii} region luminosity functions \citep[e.g.][]{Bradley2006} and is comparable to that of 30~Dor in the Large Magellanic Cloud [LMC; $L$(H$\alpha$)~=~3.9~$\times$~10$^{39}$~erg~s$^{-1}$ within a radius of 150~pc; \citealt{K86.ref}].

\subsection{Gas kinematics}
\label{kinematics.sec}

We further analyse the kinematics of the ionised gas using H$\alpha$ central velocity and line width. Both values are derived from Gaussian fitting to the emission lines. For the central velocity, we remove the galaxy's systematic recession velocity of 1375~km~s$^{-1}$ to highlight the internal motions within the galaxy. The line width is characterised by the standard deviation of the Gaussian function ($\sigma$) and is related to the line's full width at half maximum (FWHM) by FWHM~=~2.355$\sigma$. The observed line width is a combination of the intrinsic velocity dispersion and the effect of instrumental broadening. We use the \citet{LSF.ref} parametrisation of MUSE line spread function (LSF) and find that the instrumental broadening corresponds to 49.3~km~s$^{-1}$ at the (redshifted) wavelength of H$\alpha$. This component is removed from the observed line width using the relation $\sigma_{\rm gas}$~=~$\sqrt{\sigma_{\rm obs}^2 - \sigma_{\rm LSF}^2}$. 

Figures~\ref{im3.fig} and \ref{im4.fig} display the derived maps of H$\alpha$ central velocity and line width. The central velocity has a clear gradient from north to south, arising from the disk rotation of NGC~5806. At the position of SNe~2004dg/2012P, the ionised gas is moving toward the observer at a velocity of $\sim$25--30~km/s. Apart from this general pattern, there is a very interesting patch, $\sim$500~pc northeast to SN~2012P, which is redshifted by $\sim$20~km~s$^{-1}$ relative to its surrounding area. The line width map shows that the velocity dispersion is a few km~s$^{-1}$ in the intra-arm regions and typically 10--30~km~s$^{-1}$ along the spiral arms. Around SNe~2004dg/2012P, there is an arc of clumps with significantly higher velocity dispersions of 30--40~km~s$^{-1}$; in contrast, the velocity dispersion is only at a $\sim$25~km~s$^{-1}$ level at the SN positions.

To better investigate these kinematic structures, we overlay contours of H$\alpha$ integrated flux on the central velocity and line width maps (Fig.~\ref{turb.fig}). It is immediately obvious that the redshifted patch and the high-velocity dispersion clumps are all distributed at the outer boundary of the giant H~\textsc{ii} region. These kinematic structures could be formed as the H~\textsc{ii} region expands and shocks the ambient gas; it is also possible that the H~\textsc{ii} region triggers new episodes of star formation, which in turn injects turbulent energy into the local ISM. All these phenomena reflect the powerful feedback from the star-forming complex in the environment of SNe~2004dg/2012P.

\subsection{Region definition for local environment analysis}
\label{definition.sec}

In summary, SNe~2004dg/2012P occurred on the spiral arm of NGC~5806, where a star-forming complex is apparent and has ionised a giant H~\textsc{ii} region with its UV radiation. There is a difference between the two SNe: SN~2004dg shows a clear offset from the complex while SN~2012P is spatially coincident with a compact star cluster within the complex. For further analysis in the following sections, we define two circular regions around the two SNe (\textit{Region~A} for SN~2012P and \textit{Region~B} for SN~2004dg), which are representative of their local environments as well as their different relationships with the star-forming complex (Fig.~\ref{struct.fig}). For SN~2012P, we place the centre of \textit{Region~A} on the peak of H$\alpha$ integrated flux; with a radius of 1.25~arcsec (i.e. 162~pc), \textit{Region~A} encloses most of the star-forming complex and the core of the giant H~\textsc{ii} region. Stars around SN~2004dg are significantly fainter and redder (thus older) and more sparsely distributed than those in \textit{Region~A}. Thus, we define \textit{Region~B} with a radius of 1.0~arcsec (i.e. 130~pc) to include those sources and place its centre slightly northward of SN~2004dg to avoid overlap with \textit{Region~A}. In addition, we define a third \textit{Region~C} to the south of the star-forming complex to study the extinction along the spiral arm (Section~\ref{star_model.sec}).

\section{Analysis of the gaseous component}
\label{gas.sec}

\begin{figure*}
\centering
\includegraphics[width=0.9\linewidth, angle=0]{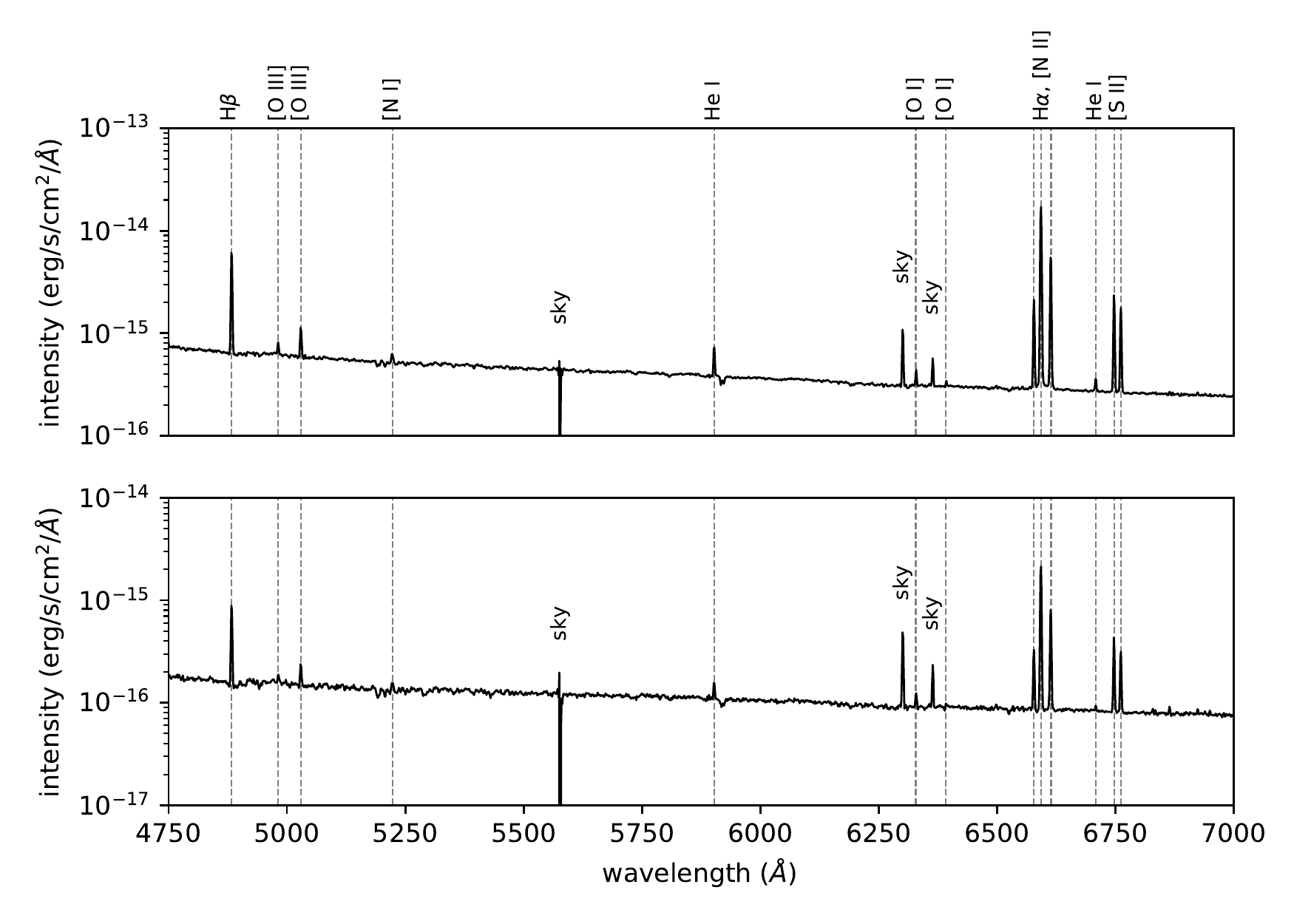}
\caption{Stacked spectra of \textit{Region~A} (top) and \textit{Region~B} (bottom). The effect of interstellar extinction has been corrected.}
\label{spec.fig}
\end{figure*}

\begin{figure*}
\centering
\includegraphics[width=0.9\linewidth, angle=0]{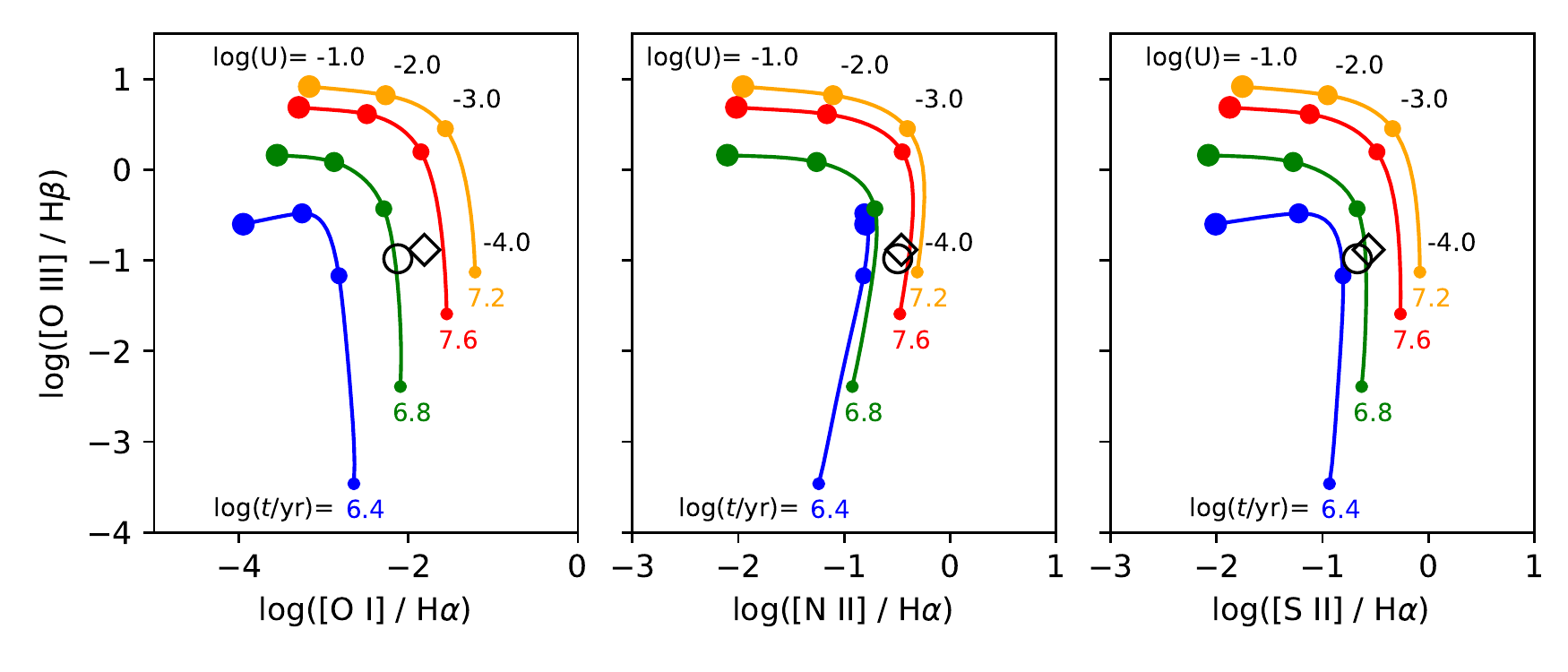}
\caption{\textit{Region~A} (open black circle) and \textit{Region~B} (open black diamond) on the BPT-like diagrams. The solid lines are for model H~\textsc{ii} regions with ages of log($t$/yr)~=~6.4 (blue), 6.8 (green), 7.2 (orange) and 7.6 (red). The ionisation parameter log($U$) varies from $-$4.0 to $-$1.0 along the lines; the filled points of different sizes show where log($U$) reaches values of $-$4.0, $-$3.0, $-$2.0 and $-$1.0. The model H~\textsc{ii} regions all have solar abundances. See the text for details.}
\label{bpt.fig}
\end{figure*}

\begin{figure}
\centering
\includegraphics[width=1.0\linewidth, angle=0]{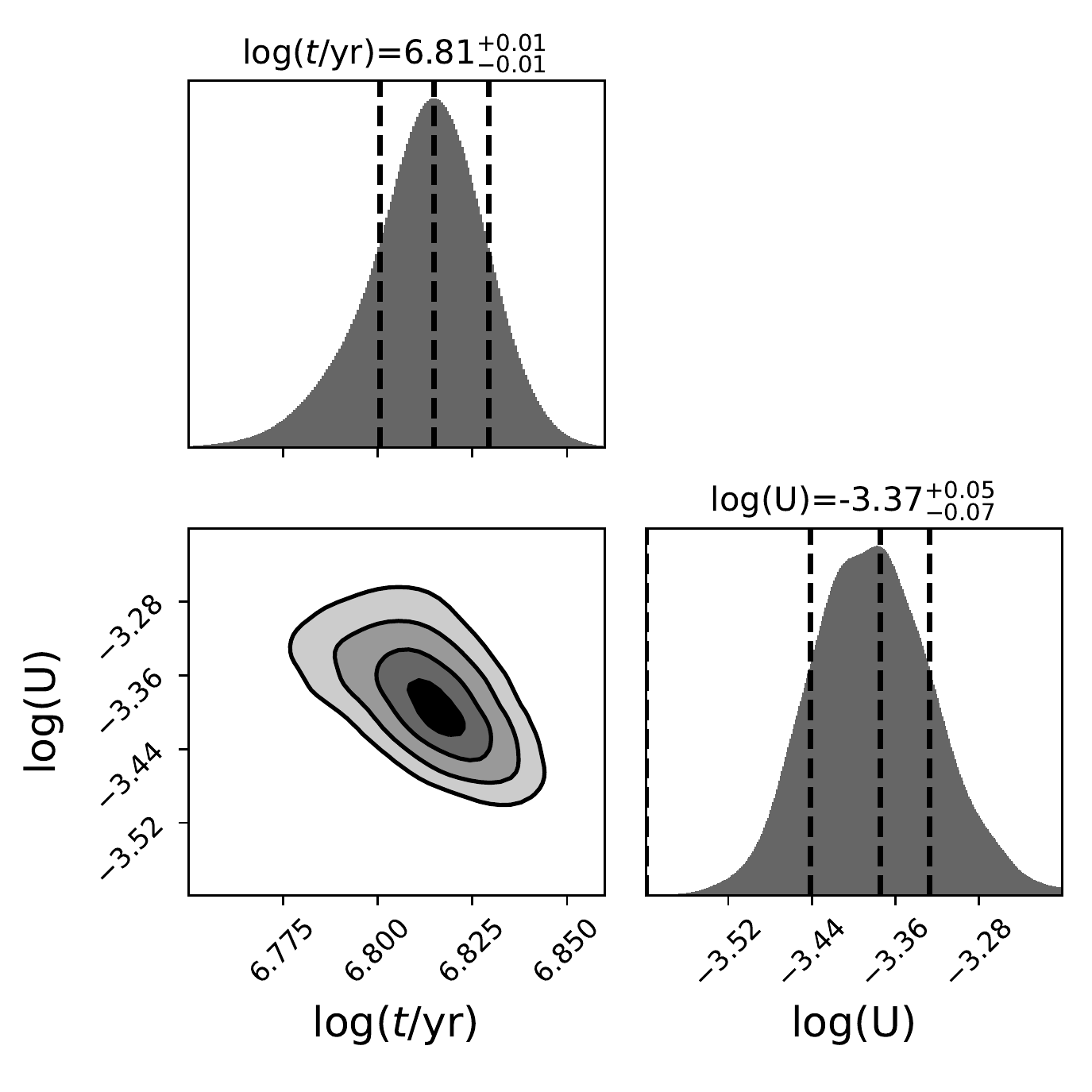}
\caption{Posterior probability distributions for the age and the ionisation parameter of the ionising stars. In the 2-dimensional plots, the contours correspond to 0.5$\sigma$,1.0$\sigma$,1.5$\sigma$ and 2.0$\sigma$ quantiles from inside to outside. In the 1-dimensional plots, the vertical lines correspond to the modes and 1$\sigma$ highest-density credible intervals.}
\label{corner.fig}
\end{figure}


\subsection{Extinction and extinction-corrected spectra}
\label{spec.sec}

Analysis of the ionised gas is based on the MUSE spectra. We first use the Balmer decrement to estimate interstellar extinction at the position of each spaxel. In doing this, we assume an intrinsic flux ratio of $I$(H$\alpha$)/$I$(H$\beta$)~=~2.87 \citep{agn2.ref} and a \citet{avlaw.ref} extinction law with $R_V$~=~$A_V$/$E(B-V)$~=~3.1. \textit{Region~B} has an extinction around $A_V$~$\sim$~1.5~mag; while part of \textit{Region~A} has a similar value, the extinction in its northern area can reach as high as $A_V$~$\sim$~2.0~mag. This is possibly due to a foreground dusty cloud to the north of SN~2012P and to the east of SN~2004dg. Note that the derived extinction is only for the ionised gas; the stars may have a different extinction if there is any dust between gas and stars along the line of sight. We use the derived extinction to correct the spectrum of each spaxel; spaxels within \textit{Region~A}/\textit{B} are then stacked together for the analysis in this section (Fig.~\ref{spec.fig}).

For both \textit{Region~A} and \textit{Region~B}, their spectra display a series of important nebular lines including the hydrogen and helium recombination lines (e.g. H$\alpha$~$\lambda$6563, H$\beta$~$\lambda$4861, He~\textsc{i}~$\lambda$5876, He~\textsc{i}~$\lambda$6678) as well as the collisionally-excited lines of some heavier elements (e.g. [O~\textsc{iii}]~$\lambda\lambda$4959, 5007; [O~\textsc{i}]~$\lambda\lambda$6300, 6363; [N~\textsc{ii}]~$\lambda\lambda$6548, 6584; [S~\textsc{ii}]~$\lambda\lambda$6717, 6731). Residuals of imperfect sky subtraction can be seen at 5578~\AA, 6300~\AA, and 6363~\AA; yet, they do not overlap with the nebular lines, which are redshifted by $z$~=~0.0045 relative to their rest wavelengths. We do not analyse the spectra at $\lambda$~$>$~7000~\AA, which suffer from large residuals of sky subtraction associated with the OH emission lines.

By fitting Gaussian functions to the emission lines, we find that \textit{Region~A} has an H$\alpha$ integrated flux of 5.18~$\times$~10$^{-14}$~erg~s$^{-1}$~cm$^{-2}$. The emission lines of \textit{Region~B} are significantly fainter than those of \textit{Region~A} by an order of magnitude; their line ratios, however, are very close to each other. Figure~\ref{bpt.fig} shows the \citet[BPT;][]{bpt.ref}-like diagrams using line ratios of [O~\textsc{iii}]~$\lambda$5007/H$\beta$, [O~\textsc{i}]~$\lambda$6300/H$\alpha$, [N~\textsc{ii}]~$\lambda$6584/H$\alpha$ and [S~\textsc{ii}]~($\lambda$6717+$\lambda$6731)/H$\alpha$ (for simplicity, we shall denote these line ratios as [O~\textsc{iii}]/H$\beta$, [O~\textsc{i}]/H$\alpha$, [N~\textsc{ii}]/H$\alpha$ and [S~\textsc{ii}]/H$\alpha$). It is clear that \textit{Region~A} and \textit{Region~B} have very similar positions on the BPT-like diagrams. This is reasonable since they sample the same H~\textsc{ii} region, one on the core while the other on the outer area. This also suggests that they should have very similar nebular properties. Thus, we focus on the spectrum of \textit{Region~A} in the following analysis.

\subsection{Metallicity, electron temperature and density}
\label{properties.sec}

We estimate the gas-phase metallicity via the strong-line diagnostics, using line ratios of [O~\textsc{iii}]~$\lambda$5007/H$\beta$ and [N~\textsc{ii}]~$\lambda$6584/H$\alpha$ and the O3N2 calibration of \citet{o3n2.ref}. An oxygen abundance of 12~+~{\rm log(O/H)} = 8.65~$\pm$~0.18~dex is derived, which is very close to the solar value \citep[8.69~dex;][]{solar.ref} and consistent with the result of \citetalias{F16} with the N2 method (8.61~$\pm$~0.18~dex).

The electron temperature of the ionised gas can be estimated with the [O~\textsc{i}]~($\lambda$6300+$\lambda$6363)/$\lambda$5577 or the [N~\textsc{ii}]~($\lambda$6548+$\lambda$6584)/$\lambda$5755 ratio. Unfortunately, neither [O~\textsc{i}]~$\lambda$5577 nor [N~\textsc{ii}]~$\lambda$5755 is detected in the spectrum; as a result, we can only get an upper limit for the electron temperature, which is $T_{\rm e}$~$<$~8000~K according to the \citet{agn2.ref} calibration. Also based on their calibration, we constrain the electron density to be $n_{\rm e}$~$\lesssim$~30~cm$^{-3}$ with the [S~\textsc{ii}]~$\lambda$6716/$\lambda$6731 diagnostics.

\subsection{Modelling of the ionised gas}
\label{gas_model.sec}

We model the ionised gas as a spherical shell around the ionising stars; the ionising radiation illuminates the shell and undergoes radiative transfer through the gas. The ionising spectrum is simulated with the \textsc{bpass} binary population and spectral synthesis \citep[version 2.1;][]{bpass.ref}; for the model stellar population, we fix the metallicity to be solar and leave its age ($t$) as a free parameter. The intensity of the ionising radiation is characterised by the ionisation parameter, defined as
\begin{equation}
U = \dfrac{Q(H)}{4 \pi r_{0}^{2} n_{\rm H} c},
\end{equation}
where $Q(H)$ is the luminosity of hydrogen-ionising photons, $r_0$ the inner radius of the gas shell, $n_{\rm H}$ the hydrogen density (i.e. the density of hydrogen atoms in all forms of particles including neutral atoms, ions, molecules, etc.) and $c$ the speed of light. The ionisation parameter is also left as a free parameter.

For simplicity, the gas shell is assumed to be dust-free and radiation-bounded (i.e. optically thick); the validity of these two assumptions will be discussed later. Consistent with the derived properties (Section~\ref{properties.sec}), we assume the gas shell has a hydrogen density of $n_{\rm H}$~=~10~cm$^{-3}$ and a solar chemical composition. Since this work uses lines of three metal elements (nitrogen, oxygen and sulphur), we introduce three parameters, [N/H], [O/H] and [S/H] (i.e. their abundances relative to solar), to account for their abundance variations. Abundances of other elements are fixed.

We then calculate the nebular spectrum with the \textsc{cloudy} package \citep[version 17.02][]{cloudy.ref}. The emission lines have different strengths for different parameters of $t$, $U$, [N/H], [O/H] and [S/H] (as an example, Fig.~\ref{bpt.fig} shows the predicted line ratios on the BPT-like diagrams). Thus, one can infer the properties of the ionised gas and the ionising stars by fitting the model to the observed spectrum.


\newcommand*{\gasLogA}{\ensuremath{6.81^{+0.01}_{-0.01}}}
\newcommand*{\gasLogAmode}{\ensuremath{6.81}}
\newcommand*{\gasLogAerr}{\ensuremath{0.01}}
\newcommand*{\gasLogU}{\ensuremath{-3.37^{+0.05}_{-0.07}}}
\newcommand*{\gasLogUmode}{\ensuremath{-3.37}}\newcommand*{\gasLogUerr}{\ensuremath{+0.05/-0.07}}
\newcommand*{\gasMetN}{\ensuremath{0.2~\pm~0.1}}
\newcommand*{\gasMetNmode}{0.2}
\newcommand*{\gasMetNerr}{0.1}
\newcommand*{\gasMetO}{\ensuremath{0.0~\pm~0.1}}
\newcommand*{\gasMetOmode}{0.0}
\newcommand*{\gasMetOerr}{0.1}
\newcommand*{\gasMetS}{\ensuremath{-0.1~\pm~0.1}}
\newcommand*{\gasMetSmode}{\ensuremath{-0.1}}
\newcommand*{\gasMetSerr}{0.1}

We use four important line ratios to compare the model with observations, i.e. [O~\textsc{iii}]/H$\beta$, [O~\textsc{i}]/H$\alpha$, [N~\textsc{ii}]/H$\alpha$ and [S~\textsc{ii}]/H$\alpha$. These lines are strong with high SNRs and their fluxes are less affected by stellar absorption; lines of each pair have similar wavelengths, so the line ratios are insensitive to uncertainties in extinction correction or flux calibration. The likelihood function is defined as
\begin{equation}
L = \prod_{j} \dfrac{1}{\sqrt{2\pi \left(\sigma_{j, \rm mod}^2 + \sigma_{j, \rm obs}^2 \right)}} \times {\rm exp} \left[ - \dfrac{1}{2} \dfrac{\left(r_{j, \rm mod} - r_{j, \rm obs} \right)^2}{\sigma_{j, \rm mod}^2 + \sigma_{j, \rm obs}^2}  \right],
\end{equation}
where $r_{j, \rm mod}$ ($r_{j, \rm obs}$) is the logarithm of the $j$th line ratio from model (observation) and $\sigma_{j, \rm mod}$ ($\sigma_{j, \rm obs}$) is the uncertainty of $r_{j, \rm mod}$ ($r_{j, \rm obs}$). The value of $\sigma_{j, \rm obs}$ is taken from the line fitting error and $\sigma_{j, \rm mod}$ is based on the assumption that the models have a 10\% uncertainty in the line flux. We use logarithmic line ratios to avoid the likelihood function being dominated by one ratio that is much larger than the other ones. Flat priors are used for log($t$/yr) and log($U$). For [N/H], [O/H] and [S/H], we assume Gaussian priors centred on zero with a standard deviation of 0.18~dex (consistent with the estimate and uncertainty of the strong-line method; Section~\ref{properties.sec}). The posterior probability distributions are solved numerically with the dynamic nested sampling package \textsc{dynesty} \citep{dynesty.ref}.

\begin{table}
\center
\caption{Fitting results for the ionised gas and the ionising stars (based on the stacked spectrum of Region~A). The quoted values are the modes and 1$\sigma$ highest-density credible intervals of the posterior probability distributions.}
\begin{tabular}{cccc}
\hline
\hline
Parameter & Value & Error & Note \\
\hline
log($t$/yr) & \gasLogAmode & \gasLogAerr & age of the ionising stars \\
log($U$) & \gasLogUmode & \gasLogUerr & ionisation parameter \\
${\rm [N/H]}$ & \gasMetNmode & \gasMetNerr & nitrogen abundance\\
${\rm [O/H]}$ & \gasMetOmode & \gasMetOerr& oxygen abundance \\
${\rm [S/H]}$ & \gasMetSmode & \gasMetSerr & sulphur abundance \\
\hline
\end{tabular}
\label{gas.tab}
\end{table}

The fitting results are provided in Fig.~\ref{corner.fig} and Table~\ref{gas.tab}. The ionising stars have a very young age of only log($t$/yr)~=~\gasLogA\ and a low ionisation parameter of log($U$)~=~\gasLogU. Abundances of oxygen and sulphur are still consistent with solar (i.e. zero) within their 1$\sigma$ credible intervals. In contrast, the nitrogen abundance is slightly more enriched by [N/H]~=~\gasMetN~dex (i.e. by a scaling factor of $\sim$1.5). This corresponds to a nitrogen-to-oxygen abundance ratio of log(N/O)~=~$-$0.7, adopting the absolute scale for solar abundances by \citet{solar.ref}. We note that the nitrogen abundance is still within the observed range for H~\textsc{ii} regions \citep[e.g.][]{Pilyugin2003}. It is necessary to consider the abundance variations; if we assume all elements to have exactly solar abundances, the nitrogen lines of the best-fitting model will be too weak compared with the observations.

In the above analysis, we have ignored the effect of stellar absorption when measuring the emission line fluxes. We suggest that this effect is very small for the spectrum of \textit{Region~A}. For instance, we measured an equivalent width (EW) of 181~\AA\ for H$\alpha$ and 28~\AA\ for H$\beta$. For a stellar population of log($t$/yr)~=~6.81, which dominates the spectrum of \textit{Region~A}, the stellar absorption corresponds to an EW of 1.3~\AA\ for H$\beta$ and close to 0 for H$\alpha$ as estimated from the \textsc{bpass} models (it will be shown later that \textit{Region~A} contains two more populations, but their contributions to the total spectrum are smaller; otherwise, the age derived in this section would be different). Thus, the effect of stellar absorption is less than 5\%, smaller than the 10\% uncertainty that we have assumed for the model line fluxes.

Assuming the observed spectrum is dominated by the ionising stellar population, we compared the continuum spectrum with the \textsc{bpass} theoretical SED of log($t$)~=~6.81. An initial mass of $\sim$2~$\times$~10$^6$~$M_\odot$ is derived, corresponding to an ionising photon luminosity of $\sim$8~$\times$~10$^{51}$~s$^{-1}$. The ionising flux budget could power an H$\alpha$ luminosity of up to $\sim$10$^{40}$~erg~s$^{-1}$ and is sufficient to account for the observed value (4.4~$\times$~10$^{39}$~erg~s$^{-1}$).

\section{Analysis of the stellar components}
\label{star.sec}

\begin{figure*}
\centering
\includegraphics[width=0.95\linewidth, angle=0]{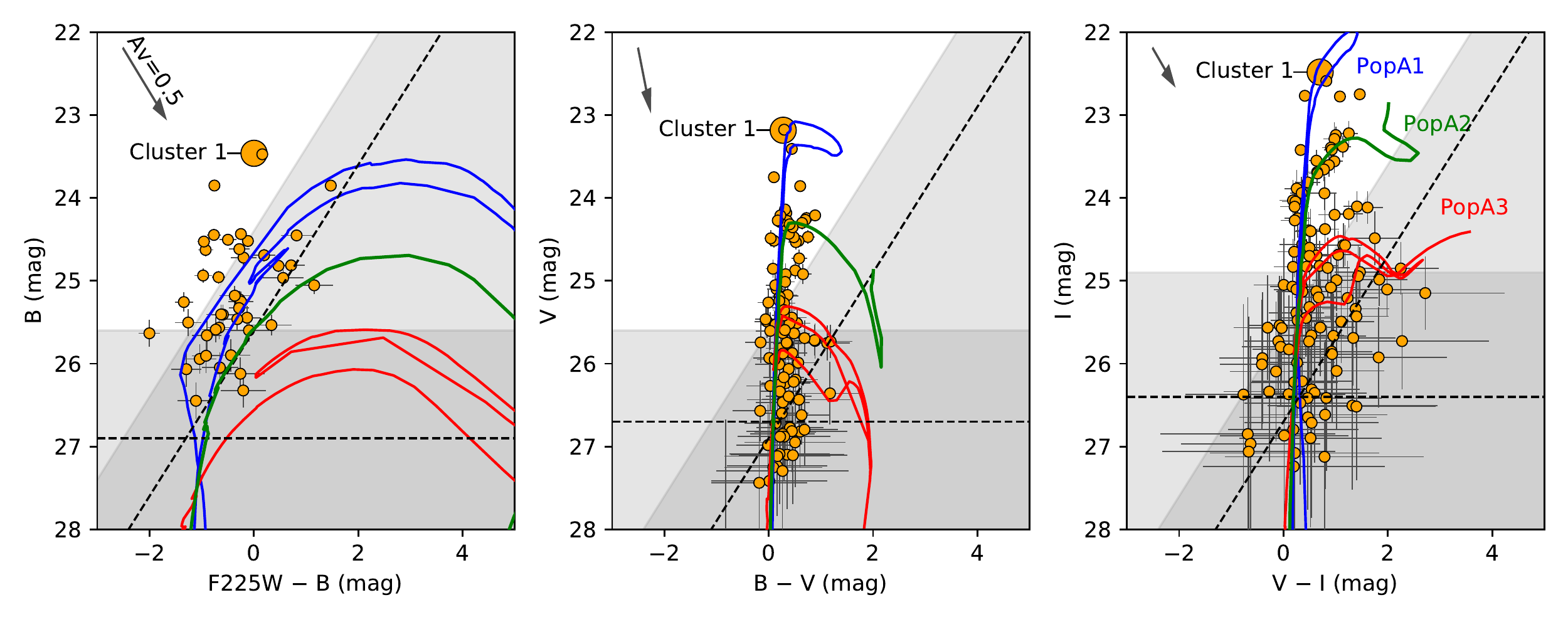}
\includegraphics[width=0.95\linewidth, angle=0]{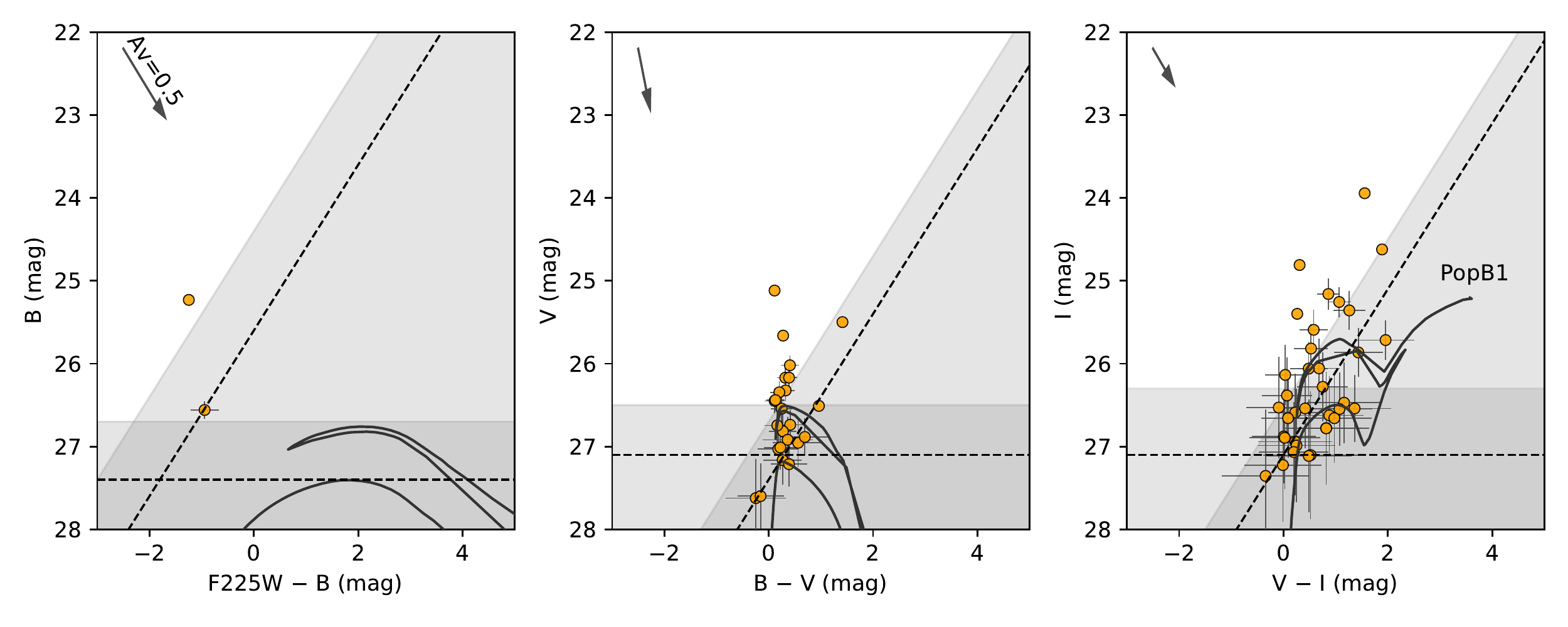}
\caption{CMDs of \textit{Region~A} (top) and \textit{Region~B} (bottom). The blue, green, red and black curves are \textsc{parsec} isochrones corresponding to the four stellar populations with different ages. The larger data point in the top row of panels corresponds to \textit{Cluster~1}, which is spatially coincident with the position of SN~2012P. The error bars reflect the photometric uncertainties. In each panel, the two dashed lines indicate the completeness level with a 50\% recovery probability of the artificial stars; the grey-shaded region shows where $\leq$~68\% of the artificial stars can be recovered. The arrow in the upper-left corner of each panel is the reddening vector of $A_V$~=~0.5~mag corresponding to a standard Galactic reddening law with $R_V$~=~3.1.}
\label{cmd.fig}
\end{figure*}


\begin{figure}
\centering
\includegraphics[width=1\linewidth, angle=0]{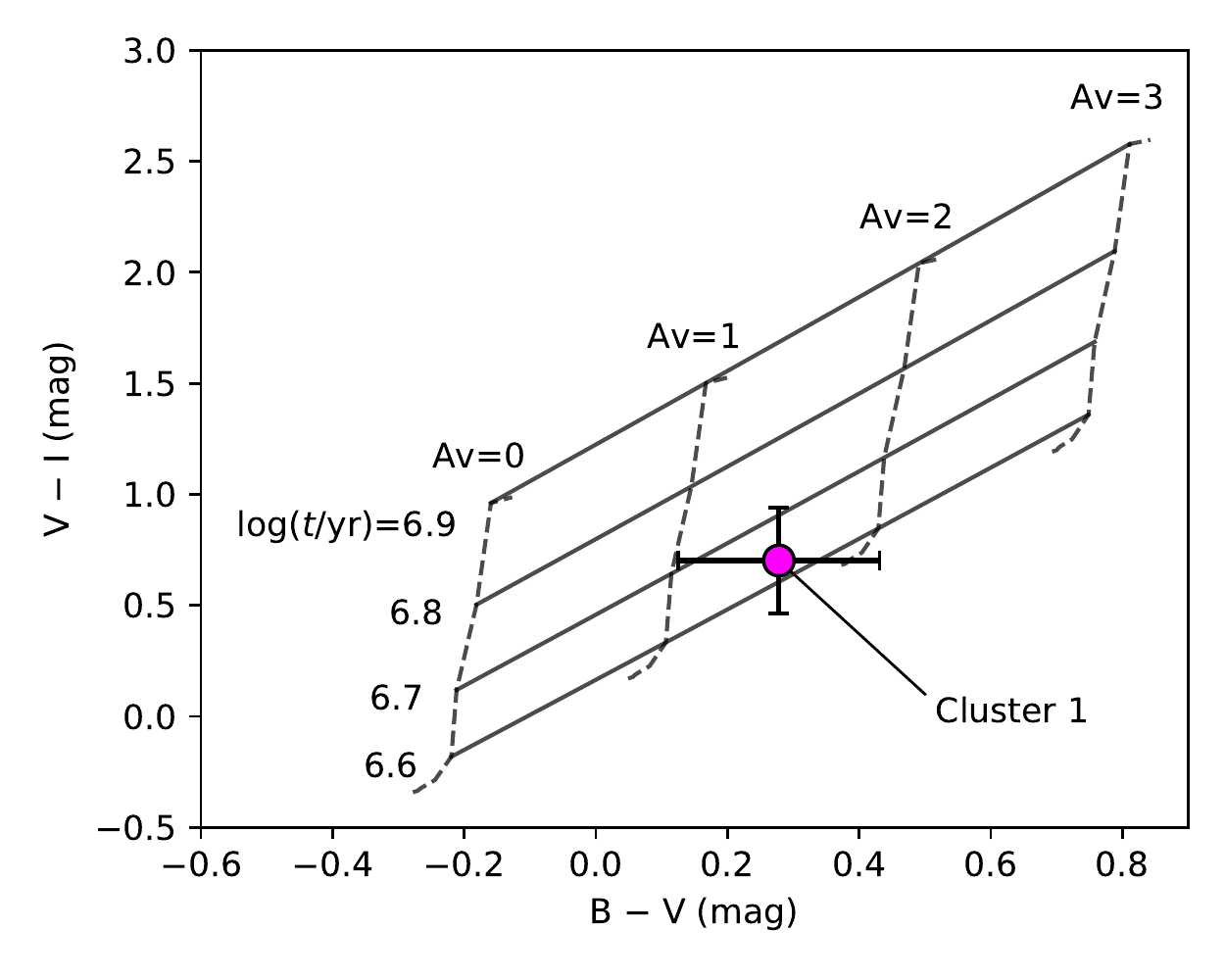}
\caption{Colour-colour diagram of \textit{Cluster~1} and theoretical predictions of \textsc{bpass} model populations. The solid lines correspond to constant ages with a varying extinction while the dashed lines have constant extinctions and a varying age. The error bars reflect 3$\sigma$ photometric uncertainties.}
\label{ccd.fig}
\end{figure}

\begin{figure}
\centering
\includegraphics[width=1.0\linewidth, angle=0]{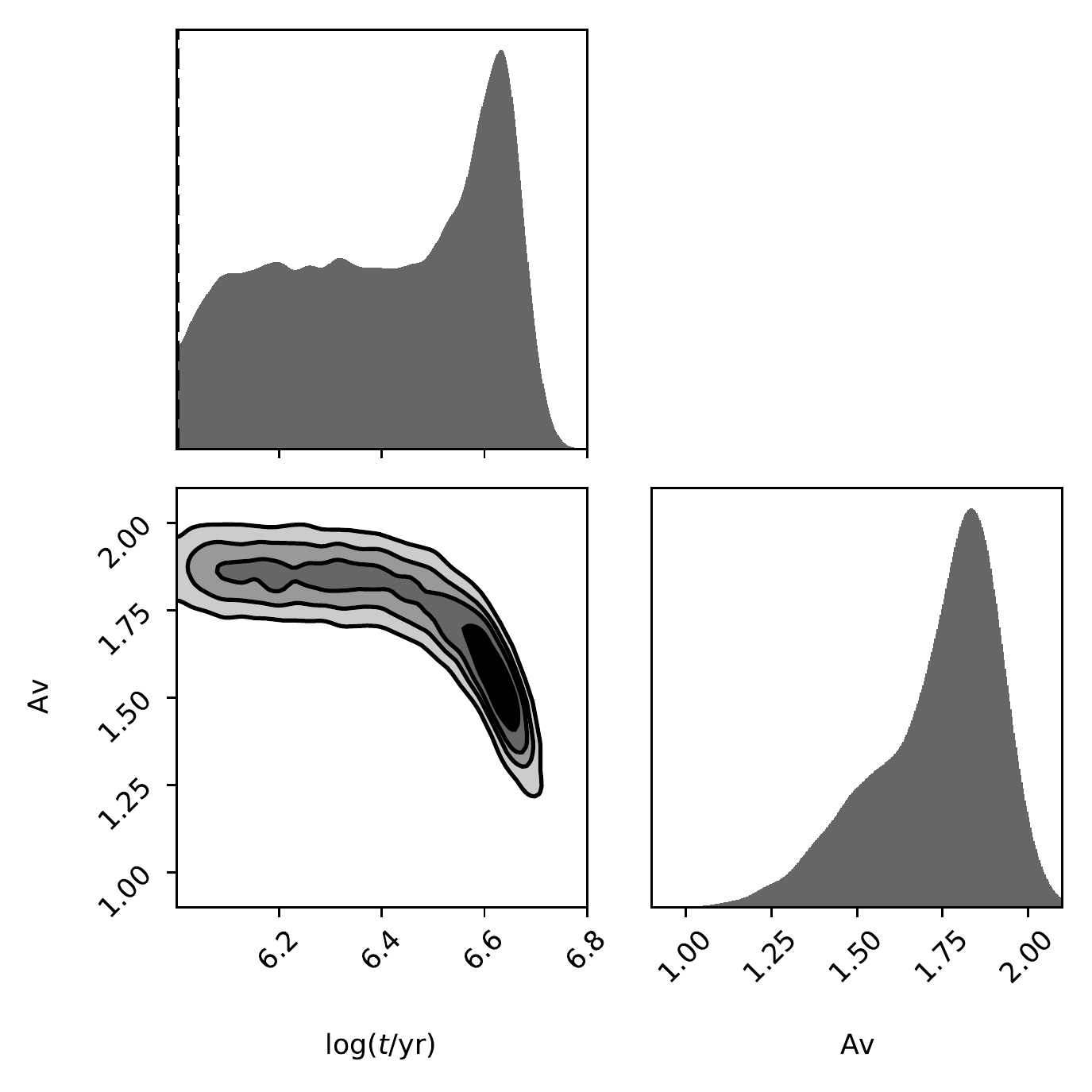}
\caption{Marginalised posterior probability distributions for \textit{Cluster~1}'s age and extinction, derived with no prior constraint on its extinction. In the 2-dimensional plots, the contours correspond to 0.5$\sigma$,1.0$\sigma$,1.5$\sigma$ and 2.0$\sigma$ quantiles from inside to outside.}
\label{nop.fig}
\end{figure}

\begin{figure}
\centering
\includegraphics[width=1.0\linewidth, angle=0]{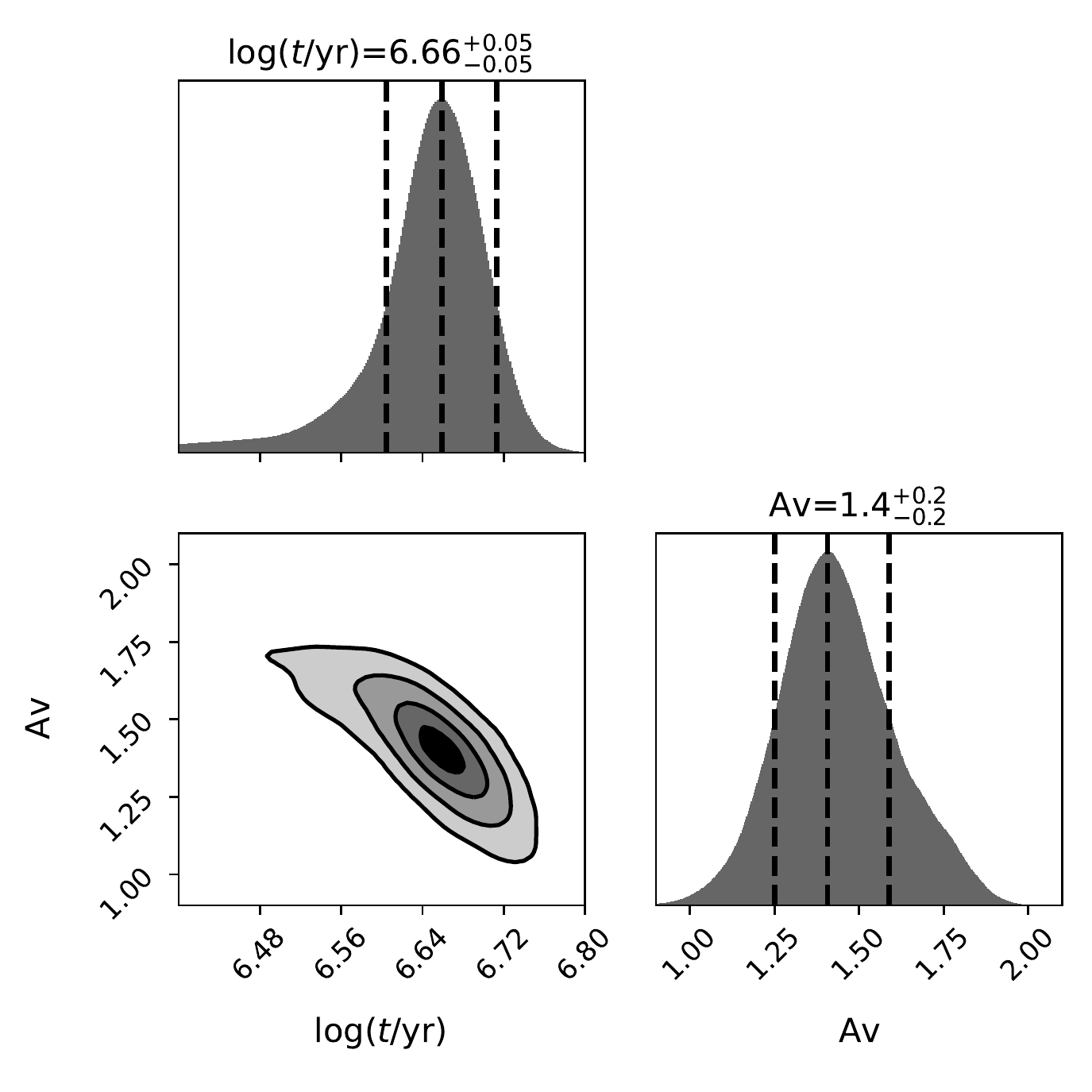}
\caption{Marginalised posterior probability distributions for \textit{Cluster~1}'s age and extinction, derived with a Gaussian prior contraint on its extinction. In the 2-dimensional plots, the contours correspond to 0.5$\sigma$,1.0$\sigma$,1.5$\sigma$ and 2.0$\sigma$ quantiles from inside to outside. In the 1-dimensional plots, the vertical lines correspond to the modes and the 1$\sigma$ highest-density credible intervals.}
\label{cluster.fig}
\end{figure}

\begin{figure}
\centering
\includegraphics[width=1.0\linewidth, angle=0]{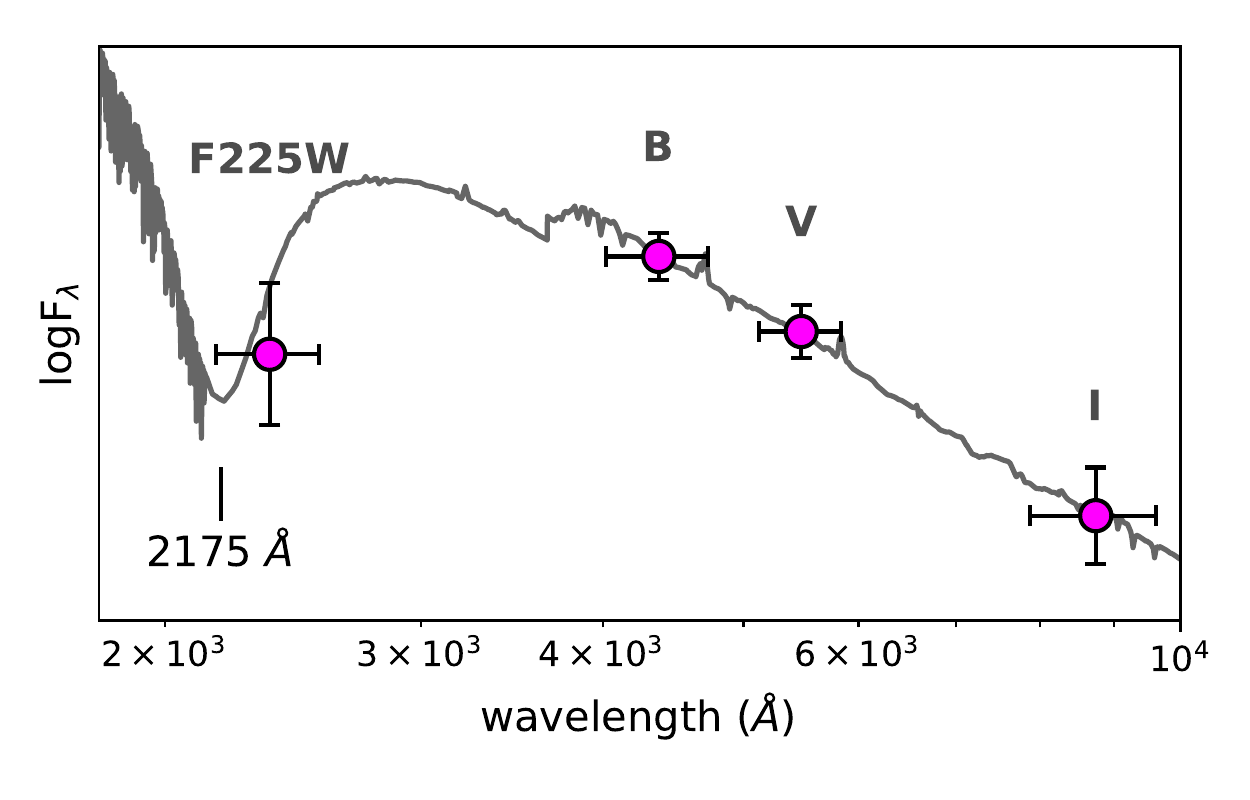}
\caption{Comparison of the observed SED of \textit{Cluster~1} (data points) and the best-fitting \textsc{bpass} model spectrum. The vertical error bars reflect 3$\sigma$ photometric uncertainties and the horizontal error bars correspond to the bandwidths. The vertical axis is in arbitrary units. The dip in the model spectrum is caused by the 2175~\AA\ bump in the adopted Galactic extinction law.}
\label{sed.fig}
\end{figure}

\subsection{Photometry}
\label{phot.sec}

We further try to analyse the stellar components in the SN environments. To do this, we first run \textsc{dolphot} \citep{dolphot.ref} to derive photometry on the HST images. For each filter, we only use sources with SNR~$>$~5; all these sources are point-like, according to their shape parameters (\texttt{sharpness}) reported by \textsc{dolphot}. The spatial distribution of the detected sources are displayed in Fig.~\ref{struct.fig}. For \textit{Region~A}/\textit{B} and for each filter, we use artificial star tests to determine the completeness (defined as the magnitude at which 50\% of the artificial stars can be recovered), its ``uncertainty" (the magnitude range where the recovery rate declines from 68\% to 32\%) and an additional scaling factor for the photometric uncertainties to account for the effect of source crowding.

The filters of ACS/F435W and WFC3/F438W, as well as ACS/F555W and WFC3/F555W, are very similar to each other. Thus, magnitudes in these bands are converted to the standard B- and V-band magnitudes \citep{acs.ref, wfc3.ref} and then combined together to increase the SNRs. Slight offsets are found between magnitudes converted from different observations. These offsets may result from imperfect photometry; similar offsets have also been found in a number of studies with HST data (e.g. \citealt{Eldridge2015}; \citealt{S20a}). We apply an empirical correction by comparing the source colours with theoretical stellar loci, the details of which can be found in Appendix~\ref{offset.sec}. For consistency, we also convert the ACS/F814W magnitudes to the standard I-band magnitudes \citep[these two filters are very similar and the correction is very small;][]{acs.ref}. The WFC3/F225W magnitudes are left unchanged in the native filter. In summary, we obtain two point-source catalogues, for \textit{Region~A} and \textit{Region~B}, in the F225W and BVI bands.

\subsection{Colour-magnitude diagrams}
\label{cmd.sec}

The top row of Fig.~\ref{cmd.fig} shows \textit{Region~A}'s sources on the CMDs. For comparison, we overlay a \textsc{parsec} \citep[version 1.2S;][]{parsec.ref} isochrone with a solar metallicity and an age of log($t$/yr)~=~\gasLogAmode\ (green curve) as estimated by modelling the ionised gas (Section~\ref{gas_model.sec}). The isochrone has been shifted by NGC~5806's distance modulus and reddened by $A_V$~=~1.0~mag with a Galactic extinction law with $R_V$~=~3.1 \citep{avlaw.ref}. This extinction value is found by visual inspection of the (B~$-$~V, V) and (V~$-$~I, I) CMDs (the F225W band is not considered since any possible variation of the extinction law has a much larger effect on this band than on the optical bands).

In the optical CMDs, the stellar isochrone of log($t$/yr)~=~\gasLogAmode\ agrees very well with the data points. In particular, the isochrone matches the tip of the main sequence at V~=~24.3~mag in the (B~$-$~V, V) diagram and the upper part of the main sequence with I~$>$~24.0~mag in the (V~$-$~I, I) diagram. Thus, our modelling of the ionised gas (Section~\ref{gas_model.sec}) has derived a very reliable age for the ionising stars. A few sources appear to be significantly brighter than the isochrone; they might be compact star clusters (given the distance it is difficult to distinguish stars and clusters based on their spatial extents), unresolved binaries/multiples, rejuvenated stars, binary merger products, or massive stars from an even younger population. A number of sources can be found to the red side of the isochrone, which could be stars with higher extinctions and/or stars from older stellar populations.

In the UV-optical (F225W~$-$~B, B) diagram, the data points appear to be younger than the stellar isochrone of log($t$)~=~\gasLogAmode. They may belong to a younger stellar population, which, however, remains inconclusive. Note that the F225W band suffers from a higher extinction than the optical bands; as a result, any dispersion in the extinction or variation of the extinction law will have a large effect on the UV magnitudes. In particular, the F225W band contains the 2175~{\AA} bump in the extinction law, which may be different or even not exist in different environments \citep[e.g.][]{smcavlaw.ref}.

The bottom row of Fig.~\ref{cmd.fig} shows the CMDs of \textit{Region~B}. It is clear that the sources are much fewer and fainter than those in \textit{Region~A}. Thus, stars in this region are likely to be older and more sparsely distributed.

\subsection{Modelling of the unresolved star cluster}
\label{cluster_model.sec}

\newcommand*{\clusterLogA}{\ensuremath{6.66~\pm~0.05}}
\newcommand*{\clusterLogAmode}{\ensuremath{6.66}}
\newcommand*{\clusterLogAerr}{\ensuremath{0.05}}
\newcommand*{\clusterAv}{\ensuremath{1.4~\pm~0.2}}
\newcommand*{\clusterAvmode}{\ensuremath{1.4}}
\newcommand*{\clusterAverr}{\ensuremath{0.2}}
\newcommand*{\clusterLogM}{\ensuremath{4.3~\pm~0.1}}
\newcommand*{\clusterLogMmode}{\ensuremath{4.3}}
\newcommand*{\clusterLogMerr}{\ensuremath{0.1}}

As mentioned in Section~\ref{components.sec}, SN~2012P is spatially coincident with a point source in the star-forming complex. Its brightness is much brighter than most stars in the surrounding area (Fig.~\ref{cmd.fig}) and remains unchanged before and after the explosion of SN~2012P \citep{F16}. Thus, this point source is most likely to be a star cluster, which may host SN~2012P's progenitor or just reside in chance alignment with the SN position. We shall refer to this source as \textit{Cluster~1} hereafter. The properties of \textit{Cluster~1} can be inferred by comparing its SED with theoretical predictions for single-age model stellar populations. To do this, we use the \textsc{bpass} model with interacting binaries.

Figure~\ref{ccd.fig} shows the (B~$-$~V, V~$-$~I) diagram of \textit{Cluster~1} and theoretical colours for different cluster ages and extinctions. A Galactic extinction law has been applied with $R_V$~=~3.1 \citep{avlaw.ref}. We do not use colours involving the F225W band, which may be significantly affected by the uncertain extinction law (in particular, its 2175~\AA\ bump). Comparing with the theoretical colours, it is clear that \textit{Cluster~1} has a very young age of log($t$/yr)~$\lesssim$~6.7; the extinction could be $A_V$~=~1--2~mag, considering the photometric uncertainties.

\begin{table}
\center
\caption{SED fitting results for \textit{Cluster~1}. The quoted values are the modes and 1$\sigma$ highest-density credible intervals of the posterior probability distributions.}
\begin{tabular}{cccc}
\hline
\hline
Parameter & Value & Error & Note \\
\hline
log($t$/yr) & \clusterLogAmode & \clusterLogAerr & age \\
$A_V$/mag & \clusterAvmode & \clusterAverr & extinction \\
log($M$/$M_\odot$) & \clusterLogMmode & \clusterLogMerr & initial mass \\
\hline
\end{tabular}
\label{cluster.tab}
\end{table}

We further try to fit model SEDs to the observed one; the F225W band is not used for the same reason as discussed above. We first use flat priors for the cluster's log-initial mass, log-age and extinction and the derived posterior probability distributions are shown in Fig.~\ref{nop.fig}. There is a significant degeneracy between age and extinction; this is because the theoretical locus of log($t$/yr)~$<$~6.6 is almost parallel to the reddening vector in the colour-colour diagram such that younger/older clusters may have a very similar SED if they have larger/smaller extinctions. Yet, the cluster is most likely to have an age of log($t$/yr)~$\sim$~6.6 with a lower extinction. Also note that the resolved stars in the star-forming complex has a mean extinction of $A_V$~=~1.0~mag with a dispersion of ~0.3~mag (Section~\ref{cmd.sec}); it is reasonable to assume \textit{Cluster~1} to have the same \textit{extinction probability distribution} as the resolved stars since they are all members of the same star-forming complex (note, however, that \textit{Cluster~1}'s extinction may not be the same as the resolved stars' \textit{mean extinction}). Although we cannot entirely rule out a very young age with a very high extinction, we suggest that \textit{Cluster~1} is more likely to have an age of log($t$/yr)~$\sim$~6.6 with a lower extinction.



To break the age-extinction degeneracy, we apply a Gaussian prior constraint for \textit{Cluster~1}'s extinction assuming it follows the same extinction probability distribution as the resolved stars in the star-forming complex. Fig.~\ref{cluster.fig} displays the updated marginalised posterior probability distributions. The age-extinction degeneracy is much reduced and the 1-dimensional distributions are single-peaked for both parameters. The most probable age is log($t$/yr)~=~\clusterLogA, which is younger than that derived by modelling the ionised gas (log($t$/yr)~=~\gasLogA; Section~\ref{gas_model.sec}). Table~\ref{cluster.tab} lists the fitted cluster parameters and Fig.~\ref{sed.fig} compares the observed SED with the best-fitting model spectrum.

\citetalias{F16} also analysed this star cluster and derived an age of 5.5~Myr [i.e. log($t$/yr)~=~6.74]. This value is slightly younger than our result but still consistent within 2$\sigma$ uncertainties. We note that they have applied an extinction estimated from SN~2012P's colour excess [$E(B-V)$~=~0.29$^{+0.08}_{-0.05}$~mag, or $A_V$~=~0.9~$\pm$~0.2~mag]. It will be shown later, however, that \textit{Cluster~1} and SN~2012P are likely in chance alignment and may have different extinctions.

\subsection{Modelling of the resolved stars}
\label{star_model.sec}

\newcommand*{\popAoneLogA}{\ensuremath{6.63~\pm~0.01}}
\newcommand*{\popAoneLogAmode}{\ensuremath{6.63}}
\newcommand*{\popAoneLogAerr}{\ensuremath{0.01}}
\newcommand*{\popAtwoLogA}{\ensuremath{6.81^{+0.02}_{-0.01}}}
\newcommand*{\popAtwoLogAmode}{\ensuremath{6.81}}
\newcommand*{\popAtwoLogAerr}{\ensuremath{+0.02/-0.01}}
\newcommand*{\popAthrLogA}{\ensuremath{7.13^{+0.04}_{-0.07}}}
\newcommand*{\popAthrLogAmode}{\ensuremath{7.13}}
\newcommand*{\popAthrLogAerr}{\ensuremath{+0.04/-0.07}}
\newcommand*{\popAoneAv}{\ensuremath{1.15^{+0.06}_{-0.05}}}
\newcommand*{\popAoneAvmode}{\ensuremath{1.15}}
\newcommand*{\popAoneAverr}{\ensuremath{+0.06/-0.05}}
\newcommand*{\popAtwoAv}{\ensuremath{1.15^{+0.06}_{-0.05}}}
\newcommand*{\popAtwoAvmode}{\ensuremath{1.15}}
\newcommand*{\popAtwoAverr}{\ensuremath{+0.06/-0.05}}
\newcommand*{\popAthrAv}{\ensuremath{0.65^{+0.06}_{-0.07}}}
\newcommand*{\popAthrAvmode}{\ensuremath{0.65}}
\newcommand*{\popAthrAverr}{\ensuremath{+0.06/-0.07}}

\newcommand*{\popBoneLogA}{\ensuremath{7.42~\pm~0.02}}
\newcommand*{\popBoneLogAmode}{\ensuremath{7.42}}
\newcommand*{\popBoneLogAerr}{\ensuremath{0.02}}
\newcommand*{\popBoneAv}{\ensuremath{0.62~\pm~0.08}}
\newcommand*{\popBoneAvmode}{\ensuremath{0.62}}
\newcommand*{\popBoneAverr}{\ensuremath{0.08}}

\begin{table*}
\center
\caption{Fitting results for the resolved stellar populations. The quoted values are the modes and 1$\sigma$ highest-density credible intervals of the posterior probability distributions.}
\begin{tabular}{ccccc}
\hline
\hline
Hyper-parameter & Hyper-prior & Value & Error & Note \\
\hline
log($t^{\rm B1}$/yr) & flat & \popBoneLogAmode & \popBoneLogAerr & mean log-age of \textit{PopB1} \\
$A_V^{\rm B1}$/mag & flat & \popBoneAvmode & \popBoneAverr & mean extinction of \textit{PopB1} \\
log($t^{\rm A1}$/yr) & from the modelling of \textit{Cluster~1} & \popAoneLogAmode & \popAoneLogAerr & mean log-age of \textit{PopA1} \\
$A_V^{\rm A1}$/mag & from matching the stellar isochrone & \popAoneAvmode & \popAoneAverr & mean extinction of \textit{PopA1} \\
log($t^{\rm A2}$/yr) & from the modelling of the ionised gas & \popAtwoLogAmode & \popAtwoLogAerr & mean log-age of \textit{PopA2} \\
$A_V^{\rm A2}$/mag & same as $A_V^{\rm A1}$/mag & \popAtwoAvmode & \popAtwoAverr & mean extinction of \textit{PopA2} \\
log($t^{\rm A3}$/yr) & flat & \popAthrLogAmode & \popAthrLogAerr & mean log-age of \textit{PopA3} \\
$A_V^{\rm A3}$/mag & same as $A_V^{\rm B1}$/mag & \popAthrAvmode & \popAthrAverr & mean extinction of \textit{PopA3} \\
\hline
\end{tabular}
\label{pop.tab}
\end{table*}

Next we try to derive ages and extinctions for the resolved stars in \textit{Region~A} and \textit{Region~B}. Compared with unresolved star clusters, their analysis is more complicated since they may be formed at different times and/or have different extinctions. To overcome this challenge, we use a Bayesian approach to fit model stellar populations to the observed stars. The method is based on \citet{Maund2016} with minor adaption; a detailed description can be found in Appendix~\ref{fittingresolved.sec}.

We apply this technique to the observed stars in \textit{Region~A} and \textit{Region~B}. The method assumes the fitted sources to be either single stars or non-interacting binaries. In practice, the observed sources may also be compact star clusters, interacting binaries and/or their products (e.g. rejuvenated stars). For instance, the few sources in \textit{Region~A} that lie above the green isochrone in Fig.~\ref{cmd.fig} have magnitudes comparable to that of \textit{Cluster~1}, and it is difficult to tell whether they are star clusters or individual stars; it is also difficult to distinguish whether they are massive stars from a younger stellar population or rejuvenated stars with the same age as the green isochrone. Given their uncertain object types, we discard them from the fitting (we note that there may be interacting binaries and/or very low-mass clusters that mix with the other stars in the CMDs, but it is very difficult to distinguish them based on their colours and magnitudes). Similarly, four sources in \textit{Region~B} (with $V$~$<$~26.0~mag and/or $I$~$<$~25.0~mag) are significantly brighter than the other majority of stars in this area. Their object types remain uncertain (e.g. relatively low-mass clusters, interacting binaries, etc.) and they are excluded from the fitting; although these sources could also be genuine single stars and/or non-interacting binaries from a younger stellar population, we shall focus on the dominant population for the other majority of stars in this region.

For each of the remaining sources in both regions, we calculate the likelihood of its photometry for every given stellar age and extinction. In doing this, we use a Galactic extinction law with $R_V$~=~3.1 \citep{avlaw.ref}. The effect of extinction law variation is largest in the F225W band and much smaller in the optical bands; moreover, the F225W band contains the 2175 \AA\ bump, which is highly uncertain in different environments. Thus, we discard the F225W band in the fitting and only use the optical BVI bands.

It has been shown in \citet{Walmswell2013} and \citet{Maund2016} that a model stellar population can be considered as a burst of star formation with a small age dispersion (prolonged star formation can be considered as multiple bursts of star formation). Following \citet{Maund2016} and \citet{Maund2017, Maund2018}, we assume the stellar log-age of a model population follows a Gaussian distribution with a standard deviation of 0.05~dex. The extinction is also assumed to have a Gaussian distribution with a typical standard deviation of 0.3~mag (estimated from the width of the main sequence in the CMDs, Fig.~\ref{cmd.fig}). The mean log-age and mean extinction of each model population are ``hyper-parameters" that control the priors for the individual stars' parameters. Next, we need to define the number of model stellar populations and choose an appropriate set of ``hyper-priors" for the hyper-parameters.

\subsubsection{\textit{Region~B}}

We start with \textit{Region~B}, which contains much fewer sources and is less complicated than \textit{Region~A}. To avoid overfitting, we use only one model population (hereafter \textit{PopB1}) to fit the observed stars. Flat hyper-priors are used for its mean log-age and mean extinction. The posterior probability distributions are then solved numerically with \textsc{dynesty} for the hyper-parameters (Table~\ref{pop.tab}). The model stellar population has a mean log-age of log($t$/yr)~=~\popBoneLogA\ and a mean extinction of $A_V$~=~\popBoneAv~mag. A stellar isochrone of this age and extinction is displayed in the CMDs, which matches well with the observed stars (Fig.~\ref{cmd.fig}).

The mean extinction of \textit{PopB1} is much smaller than those of the ionised gas ($A_V$~=~1.5--2.0~mag; Section~\ref{spec.sec}) and the young star-forming complex ($A_V$~$\sim$~1.0; Section~\ref{cmd.sec}). This is reasonable since much local dust is expected in star-forming regions. We carry out a similar population fitting to stars in \textit{Region~C} (see Fig.~\ref{struct.fig}) that lies on the other side of the complex. We derive a mean log-age of log($t$/yr)~=~7.42~$\pm$~0.02 and a mean extinction of $A_V$~=~0.72~$\pm$~0.12~mag, which are roughly consistent with those of \textit{PopB1}. Thus, it is likely that the spiral arm in this area has a nearly uniform extinction in the foreground, except the star-forming complex associated with much local dust. The different extinctions suggest that the ionised gas lies in the background of the star-forming complex and the older stellar populations on the spiral arm. The star-forming complex is most likely to be in the background of the older stars, since it should be very close to the gas that it has ionised.

\subsubsection{\textit{Region~A}}

Compared with \textit{Region~B}, \textit{Region~A} contains a larger number of stars and is much more complicated. The young star-forming complex contains at least two age components,  one of log($t$/yr)~=~\clusterLogA\ traced by \textit{Cluster~1} (Section~\ref{cluster_model.sec}) and the other of log($t$/yr)~=~\gasLogA\ traced by the ionised gas (Section~\ref{gas_model.sec}; see also Sections~\ref{cmd.sec}). In addition, there are many older stars in this region as revealed by the red sources in the CMDs (Fig.~\ref{cmd.fig}). Thus, we use three model populations to fit the observed stars, which we shall refer to as \textit{PopA1}, \textit{PopA2} and \textit{PopA3} (with increasing age), respectively.

Since we have already derived ages for \textit{PopA1} and \textit{PopA2} (by analysing \textit{Cluster~1} and the ionised gas, respectively), the previous results can be used as hyper-priors for their mean log-ages. Similarly, the mean extinction has been found to be $A_V$~$\sim$~1.0~mag; (by matching the isochrone to stars in the CMDs; Section~\ref{cmd.sec}); we assume it has a typical error of 0.1~mag and apply a Gaussian hyper-prior for their mean extinction. We assume \textit{PopA1} and \textit{PopA2} to have the the same mean extinction since they are both part of the young star-forming complex.

\textit{PopA3}, however, may have a different mean extinction. In the analysis of \textit{Region~B}, we suggest the star-forming complex lies in the foreground of the ionised gas and in the background of the older stars along the spiral arm. We cannot fully exclude the possibility that \textit{PopA3} and \textit{PopA1/A2} may be mixed together; however, we suggest this is less likely since the SNe from \textit{PopA3} may have expelled the surrounding gas and prohibited any significant star formation in its local area. As previously mentioned, the older stars on the spiral arm have a very uniform extinction with little spatial variation from \textit{Region~B} to \textit{Region~C} (across \textit{Region~A}). In this scenario, \textit{PopA3} is most likely to have a mean extinction smaller than that of \textit{PopA1}/\textit{PopA2} but similar to that of \textit{PopB1}. Thus, we use the derived mean extinction of \textit{PopB1} as a hyper-prior for the mean extinction of \textit{PopA3}. A flat hyper-prior is used for its mean log-age.

The best-fitting parameters are listed in Table~\ref{pop.tab} and the corresponding isochrones are overlaid in the CMDs (Fig.~\ref{cmd.fig}). Note that \textit{PopA3} has a relatively old mean log-age of log($t$/yr)~=~\popAthrLogA; at this age, many member stars have already undergone SN explosions and we expect they have expelled any local dust in their vicinity. This is consistent with the assumption that \textit{PopA3} shares a similar extinction with \textit{PopB1} and the other older stars along the spiral arm.

In the above analysis, we have left the weights of the three model populations ($w_{\rm A1}$, $w_{\rm A2}$, $w_{\rm A3}$) as free parameters to be fitted with the data. We find $\sim$13\% of the detected stars belong to the older \textit{PopA3}. For \textit{PopA1} and \textit{PopA2}, however, it is very difficult to constrain their relative weights. They largely overlap with each other on the CMDs (Fig.~\ref{cmd.fig}) and, for a general star on the main sequence, it is not easy to distinguish which model population it belongs to. Yet, this effect does not have a significant influence on the parameters of the model populations, which are most sensitive to stars that have already evolved off the main sequence. There is no significant difference in the fitted parameters if we manually change the weights and recalculate the posterior probability distributions.

\section{Inferring the SN progenitors from their environment}
\label{infer.sec}

\subsection{A simple picture of the SN environment}
\label{pic.sec}

\begin{table*}
\center
\caption{Components in the SN environment (Column~1), their ages (or the age of the ionising stars for the ionised gas; Columns~2--3), extinctions (Columns~4--5), the SNe that they host (Column~6), the progenitors' initial masses and lifetimes (Column~7--10) from pre-explosion images (SN~2004dg) or nebular spectroscopy (SN~2012P).}
\begin{tabular}{ | ccccc | ccccc |}
\hline
\hline
Component in & log($t$/yr) & Error & $A_V$ & Error & SN & $M_{\rm ini}$ & Error & log($t$/yr) & Error \\
the environment & & (dex) & (mag) & (mag) & & ($M_\odot$) & ($M_\odot$) & & (dex) \\
\hline
ionised gas
& \gasLogAmode & \gasLogAerr & 1.5--2.0 & ... 
&  &  &  &  &  \\
\textit{Cluster~1}
&  \clusterLogAmode & \clusterLogAerr & \clusterAvmode & \clusterAverr
&  &  &  &  &  \\
\textit{PopA1}
&  \popAoneLogAmode & \popAoneLogAerr & \popAoneAvmode & \popAoneAverr
&  &  &  &  &  \\
\textit{PopA2}
&  \popAtwoLogAmode & \popAtwoLogAerr & \popAtwoAvmode & \popAtwoAverr
&  &  &  &  &  \\
\textit{PopA3}
&  \popAthrLogAmode & \popAthrLogAerr & \popAthrAvmode & \popAthrAverr
& 2012P & 15 & ... & 7.14 & ... \\
\textit{PopB1}
&  \popBoneLogAmode & \popBoneLogAerr & \popBoneAvmode & \popBoneAverr
& 2004dg & $\leq$~13 & +1 & $\geq$~7.23 & $-0.05$ \\
\hline
\end{tabular}
\label{ages.tab}
\end{table*}

\begin{figure}
\centering
\includegraphics[width=1\linewidth, angle=0]{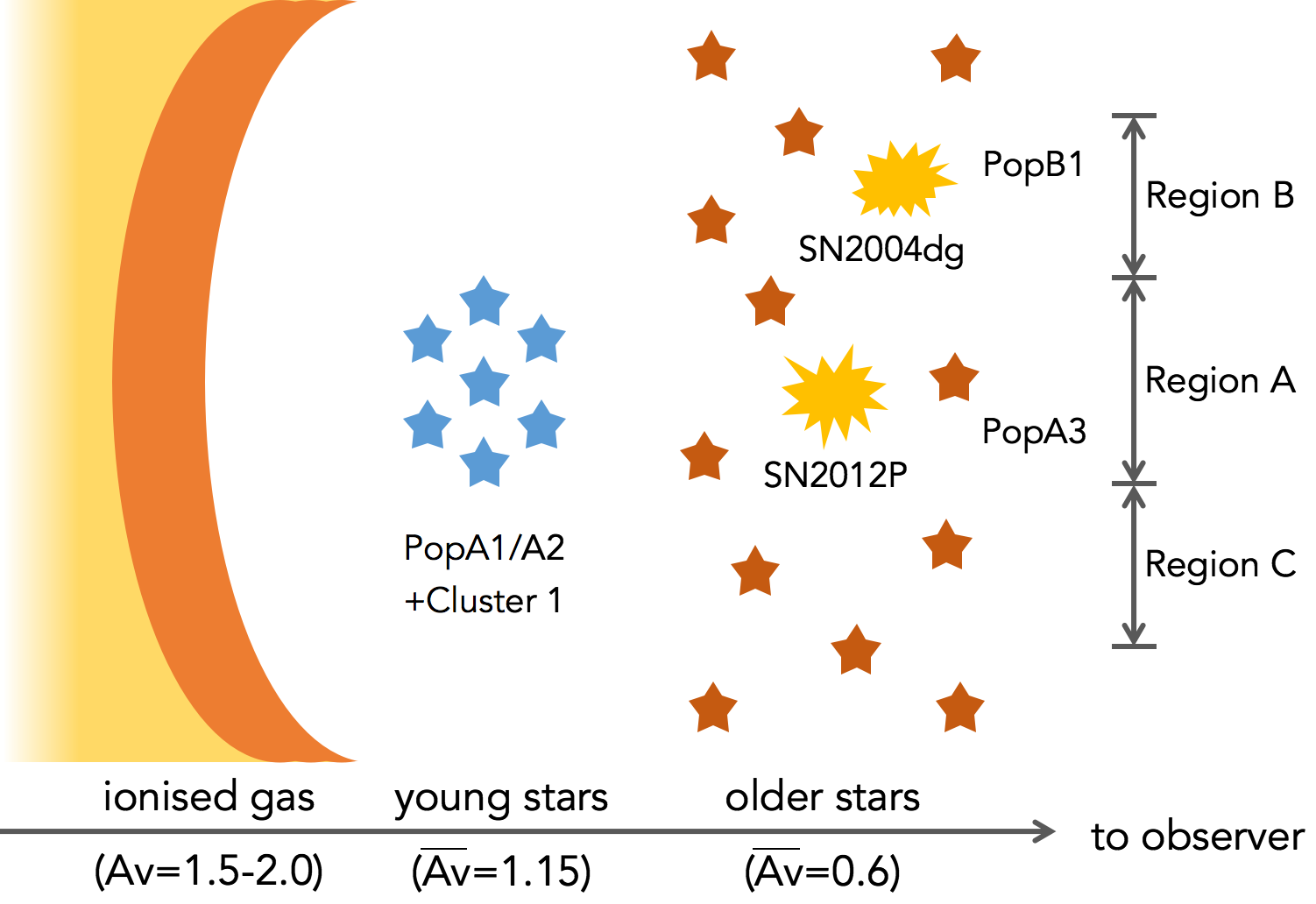}
\caption{A schematic plot of the spatial distribution of the gaseous and stellar components in the environment of SNe~2004dg/2012P along the line of sight. The younger and older stars are shown to be separate from each other; it is also possible that they are adjacent or partly overlap with each other. Dust is not displayed in the figure.}
\label{cartoon.fig}
\end{figure}

We reach a simple picture of SNe~2004dg/2012P's environment with the above detailed investigation. In summary, the most significant structures are the star-forming complex and the giant H~\textsc{ii} region. The star-forming complex is composed of two young stellar populations (\textit{PopA1} and \textit{PopA2}); in addition, we detect older stellar populations in the vicinity of both SN~2004dg (\textit{PopB1}) and SN~2012P (\textit{PopA3}). Table~\ref{ages.tab} lists the ages and extinctions derived for the various components and Figure~\ref{cartoon.fig} provides a schematic plot of the spatial distributions of the ionised gas, star-forming complex and the older stars. With different (mean) extinctions, it is most likely that the older stellar populations \textit{PopA3}/\textit{PopB1} reside in the foreground while the ionised gas is located in the background of the star-forming complex. Note that the ISM could be very patchy and the dust extinction often has high spatial variation. Thus, for an individual star, a lower (higher) extinction does not necessarily mean it is located in the foreground (background). For a population of stars, however, those in the foreground (background) statistically tend to have a lower (higher) \textit{mean} extinction.


The two SNe are located on the spiral arm of their host galaxy. As the spiral density wave propagates, the upstream low-density gas can be collected into its gravitational potential well and compressed to higher densities. Star formation can be triggered once any density fluctuations become gravitationally unstable and subject to collapse. Once formed, the young massive stars expel and ionise the surrounding gas with their powerful wind and/or ionising radiation. With time, the stars drift randomly and their spatial distributions become more and more dispersed. Finally, a star experiences a SN explosion when it reaches the end of its life. These processes occur sequentially in time but with a time lag from one to the next. As the upstream gas keeps flowing across the spiral arm's shock front, one expects a downstream sequence of gas and stars at different evolutionary stages.

The relative spatial distributions of the gas and the young/old stars may be a consequence of the above scenario [note that outside the spiral arm, the H$\alpha$ intensity and stellar density are significantly lower (Fig.~\ref{struct.fig}); thus, the observed gas and stars are likely to be physically associated rather than being in chance alignment with the spiral arm]. Similar displacements between gas and young/old stars are also commonly observed in many other spiral galaxies \citep[e.g.][]{spiral.ref}.

Gas kinematics is also relevant to the model of spiral arm-triggered star formation. The more obvious dust lanes in the eastern part of the NGC~5806's disk (Fig.~\ref{im1.fig}) suggests that its eastern (western) part is likely to be on the near (far) side; combined with the velocity map (Fig.~\ref{im3.fig}), we suggest that the disk is rotating counter-clockwise and that the spiral arms are trailing. The line-of-sight velocity at the SN position, which points toward the observer (see Section~\ref{kinematics.sec} and Fig.~\ref{im3.fig}), would be consistent with the suggested scenario if the SN position is inside the corotation radius. In this case, gas in the upstream overtakes the spiral arm from the background and then moves in the downstream toward the observer. The de-projected distance between the SN position and the galaxy's centre is $\sim$5~kpc (using inclination and position angles from \citealt{Kassin2006}). Unfortunately, there has been no reported determination of NGC~5806's co-rotation radius. We note that in a compiled catalogue of 27 galaxies by \citet{Scarano2013} only 7 galaxies with very fast spiral arm pattern speeds have corotation radii smaller than 5~kpc. It would help to test the suggested model if the co-rotation radius could be determined with future observations and analysis.

It remains unknown whether there are any dense gas clouds and/or ongoing star formation in the background of the ionised gas as expected if the spiral arm is collecting the upstream gas and triggering gravitational collapse. Radio and infrared observations are required to search for such components.

\ \\


Most stars form not in isolation but in groups of various mass and size scales. Thus, the SN progenitors are very likely to have been a member of one of the stellar populations in their environments. In particular, the progenitors' lifetimes can be estimated by the ages of their host populations. In the following we discuss which stellar populations host SNe~2004dg/2012P and what can be inferred for their progenitors. The results are then compared with those from pre-explosion images or nebular spectroscopy.

\subsection{SN~2004dg}
\label{sn2004dg.sec}


SN~2004dg is a hydrogen-rich Type~II-P SN arising from the explosion of RSGs. Its local environment (\textit{Region~B}) is dominated by a single stellar population (\textit{PopB1}). The SN progenitor would have a lifetime of log($t$/yr)~=~\popBoneLogA\ if it is a member of this population [note that SN~2004dg's extinction \citep[$A_V$~=~0.74~$\pm$~0.09~mag;][]{Smartt2009} is consistent with that of \textit{PopB1} ($A_V$~=~\popBoneAv) considering the measurement uncertainties and the extinction dispersion]. Assuming single-star evolution, the lifetime corresponds to an initial mass of $M_{\rm ini}$~=~\iipmass~$M_\odot$ acoording to the \textsc{PARSEC} (version 1.2S) stellar isochrones and evolutionary tracks \citep{parsec.ref}.

Using archival WFPC2 images, \citet{Smartt2009} did not detect the progenitor of SN~2004dg to a 3$\sigma$ magnitude limit of $m_{\rm F814W}$~=~25.0~mag. Based on the non-detection, they derived an initial mass limit of $M_{\rm ini}$~$\leq$~12~$M_\odot$ for a RSG progenitor. \citet{Davies2018} revised the bolometric correction and found a tighter constraint of $M_{\rm ini}$~$\leq$~9.5$^{+0.7}$~$M_\odot$. Both studies used a kinematic distance of 20.0~$\pm$~2.6~Mpc (i.e. $\mu$~=~31.51~$\pm$~0.13) for its host galaxy, which, however, may be affected by the galaxy's peculiar velocity. \citet{Tully2013} derived a larger distance (26.8$^{+2.6}_{-2.4}$~Mpc, $\mu$~=~32.14~$\pm$~0.20) with the Tully-Fisher relation, which should be more accurate. With the updated distance, we re-derive an initial mass limit of $M_{\rm ini}$~$\leq$~13$^{+1}$~$M_\odot$ (bolometric correction and all other parameters are the same as \citealt{Davies2018}); the quoted error arises from the distance uncertainty. Assuming single-star evolution, the non-detection corresponds to a lifetime limit of log($t$/yr)~$\geq$~7.23$_{-0.05}$. Thus, the result from environment analysis is well consistent with the progenitor non-detection on pre-explosion images.

The above analysis assumes a single-star progenitor. \citet{Z19} suggested that Type~II SNe may have very diverse progenitor channels. For instance, the progenitor may have merged with a binary companion. Alternatively, the progenitor may have accreted mass from the initially more massive companion star and then get rejected when the companion star explodes as a SN \citep{Eldridge2011, Renzo2019}. In these cases, the mass from pre-explosion images is often expected to be larger than that from the surrounding stellar population \citep{Z20}.



\subsection{SN~2012P}
\label{sn2012p.sec}

SN~2012P is a Type~IIb SN with hydrogen lines apparent only at early times. There are three stellar populations (\textit{PopA1} with \textit{Cluster~1}, \textit{PopA2}, and \textit{PopA3}) in its local environment. The progenitor would have a very short lifetime of log($t$/yr)~$<$~6.7 if it were a member of the youngest population. This corresponds to a star with an initial mass of at least $\sim$42~$M_\odot$. Such a massive star has very powerful stellar wind, which can completely strip its hydrogen envelope before the end of its life. Thus, we expect a hydrogen-free Type~Ib/c SN from it, which is in contradiction to the spectral classification of SN~2012P.

If the progenitor were coeval with \textit{PopA2}, it would have a slightly longer lifetime of log($t$/yr)~$\sim$~6.8 and an initial mass of $M_{\rm ini}$~$\sim$~31~$M_\odot$. In this case, we still expect a Type~Ib/c SN if we adopt the empirical threshold of 25~$M_\odot$ for a solar-metallicity star to become a WR star via stellar wind \citep{Crowther2006, Crowther2007}. The inferred initial mass for the progenitor is close to but slightly lower than the theoretical mass range (32.5--33~$M_\odot$) found by \citet{Claeys2011} for single Type~IIb progenitors. Several other works report different thresholds for WR stars or mass ranges for Type~IIb SN progenitors \citep[e.g.][]{Heger2003, Eldridge2004, Georgy2009, Georgy2012, Yoon2010, Groh2013, Sravan2019, Shenar2020}; the results are very sensitive to the adopted mass-loss prescription \citep{Smith2014}. Furthermore, it requires fine tuning of the initial mass to have the right stellar wind so that a residual hydrogen envelope can be left before core collapse \citep{Yoon2010, Yoon2017, Sravan2019}. Given these reasons, although we cannot fully exclude this possibility, we argue that the SN progenitor is less likely to be a member of this stellar population.


Thus, the progenitor of SN~2012P is most likely associated with the oldest \textit{PopA3} and have a lifetime of log($t$/yr)~=~\popAthrLogA\ [SN~2012P has an extinction of $A_V$~=~0.9~$\pm$~0.2~mag according to the measurement of \citetalias{F16}; this is consistent with that of \textit{PopA3} ($A_V$~=~\popAthrAv~mag) considering the measurement uncertainties and the extinction dispersion]. The lifetime corresponds to an initial mass of $M_{\rm ini}$~=~\iibmass~$M_\odot$ assuming that binary stripping has a very small effect on the lifetime of the progenitor \citep{Z17}. A star of this mass has a relatively weak stellar wind; thus, the progenitor should have been partially stripped via binary interaction, after which a low-mass hydrogen envelope was left before core collapse. Note that the binary channel is theoretically much more probable for Type~IIb SNe, without any need for fine-tuning of the initial mass and the mass loss; interestingly, three out of four binary companion detections so far have been specifically next to Type~IIb SN progenitors (SN~1993J, \citealt{Maund2004, Fox2014}; SN~2001ig, \citealt{Ryder2018}; SN~2011dh, \citealt{Folatelli2014}), supporting a binary progenitor channel for Type~IIb SNe. By independently studying the nebular-phase spectroscopy, \citetalias{F16} find SN~2012P's progenitor has an (final) oxygen mass of $\sim$0.8~$M_\odot$ and is consistent with an initial stellar mass of 15~$M_\odot$. This result is in good agreement with that from our environment analysis.



\section{Discussion}
\label{discussions.sec}

\subsection{Alignment with correlated star formation}
\label{chance.sec}

As mentioned, SN~2012P is spatially coincident with a compact star cluster (\textit{Cluster~1}) within a giant star-forming complex. Its progenitor, however, is most likely to come from an older stellar population in the foreground (Section~\ref{sn2012p.sec}). Note that the complex has a large spatial extent ($\sim$300~pc) and its member sources are very crowded. For a random star within \textit{Region~A}, we estimate a $\sim$17\% ($\sim$68\%) probability for its spatial coincidence with a young star/star cluster within 1 (2) ACS/WFC pixel(s).

On a larger scale, however, it is not really fair to say SN~2012P is in ``chance" alignment with the star-forming complex if we consider the scenario as discussed in Section~\ref{pic.sec} (see also Fig.~\ref{cartoon.fig}). Their spatial distributions may not be fully random since they were possibly formed sequentially as the spiral arm's shock front swept through their positions and triggered episodes of star formation. Such cases could be very common. \citet{Aramyan2016}, for instance, find a large number of core-collapse SNe in the downstream of spiral arms in grand-design spiral galaxies; this phenomenon is consistent with the scenario for SN~2012P.

It is therefore important to consider physical processes that may control star formation on various scales. On a galactic scale, such processes include, for example, spiral density waves (as in this case), bar perturbations \citep[e.g.][]{Ma2018}, gas accretion and infall \citep[e.g.][]{Fukui2017}, and collision/merger with other gas-rich galaxies \citep[e.g.][]{deGrijs2003}. On pc up to kpc scales, stars formed in earlier episodes can trigger new star formation in the vicinity via their radiative and mechanical feedback \citep[e.g.][]{Deharveng2005, Koenig2008}. Moreover, the ubiquitous supersonic turbulence drives the ISM hydrodynamics over a wide range of scales \citep{MacLow2004}; as a result, one often sees hierarchical patterns of star formation following the cascade of turbulent energy from large to small scales \citep[e.g.][]{Gouliermis2015, Gouliermis2017, Grasha2017a, Grasha2017b, S17a, S17b, S18}.

If we regard the observed stars as arising from a mixture of single-aged populations, each formed in a burst of star formation on small scales, these populations will appear correlated both spatially and temporally when star formation is controlled by any physical processes on larger scales. In such cases (which are very common), there is a high probability of finding younger stars in the vicinity of the SN progenitors which are formed in earlier episodes of star formation. Thus, the SN progenitors and younger stars may often reside along the same line of sight when they have the the right orientation.


This effect may partly explain the progenitor mass discrepancy between light-curve modelling and environment analysis for stripped-envelope SNe. By modelling the observed light curves, recent studies \citep{Lyman2016, Prentice2016, Taddia2018} suggest that nearly all of them have low ejecta masses consistent with moderately massive progenitors stripped in binaries. On the other hand, a large fraction of stripped-envelope SNe are spatially associated with significantly younger stars in their environments \citep[e.g.][]{Maund2016, Maund2018}. For the environment analysis, alignment with correlated star formation may lead to an overestimate of progenitor masses, if one mistakenly assume older SN progenitors are coeval with the younger stars that formed more recently. Yet, comprehensive analysis is still needed to fully resolve the discrepancy and reveal the progenitor channels of stripped-envelope SNe.

\subsection{Toward a better understanding of SN environments}
\label{toward.sec}

As demonstrated in this work, the properties of SN progenitors can be accurately and reliably inferred from their environments. However, this can only be achieved if all components in the environment are properly analysed and, in particular, their relationships with the SN progenitors are correctly identified. To this end, we have explored the combined usage of spatially-resolved IFU spectroscopy with high-spatial resolution imaging. Accordingly, sophisticated analysis techniques are needed to extract information from the combined datasets as deeply and accurately as possible.

\subsubsection{IFU spectroscopy and the ionised gas}
\label{gas_comments.sec}

IFU spectroscopy has been developed as a novel and powerful tool to investigate SN environments \citep{L11, K13a, K13b, K18, G14, G16a, G16b, G18, X19, Schady2019}. Powered by the UV radiation, the nebular emission lines from the ionised gas serve as a sensitive age indicator for the ionising stars. The emission lines are often very strong so high SNRs can be achieved with relatively short exposure times. Moreover, the commonly used nebular lines (e.g. H$\alpha$) have optical wavelengths, at which uncertainties of extinction and extinction law are less important than at UV wavelengths. In particular, IFU spectroscopy can reveal the gas properties not just at a single point, or 1-dimensionally along a slit, but also 2-dimensionally across the plane of sky.

In the modelling of the ionised gas, we have assumed a single age for the ionising stars (Section~\ref{gas_model.sec}). The HST observations, however, reveal two stellar populations of different ages in the star-forming complex (\textit{PopA1}/\textit{PopA2}; Section~\ref{star_model.sec}). Yet, the derived age is very consistent with that of \textit{PopA2}, as suggested by the CMDs and the LF (Section~\ref{cmd.sec}). This population may dominate the ionising source spectrum; it is also possible that \textit{PopA1}'s UV radiation is still confined by the very local star-forming gas.

One needs to choose an appropriate set of quantities to compare models with observations. Recent SN environment studies have extensively used the H$\alpha$ equivalent width (EW) as an age indicator. However, it may suffer from a number of uncertainties that can affect the age estimate: (1) the modelled continuum comes solely from the young ionising stars while in practice the observed continuum is contributed by stars of all ages along the same line of sight; (2) as mentioned, the observed ionised gas lies in the background of and has a higher extinction than the ionising stars; in this case, the EW is no longer extinction-independent and is smaller than expected if we ignore this effect; (3) H~\textsc{ii} regions often have covering factors (the fraction of sky covered by the gas as viewed from the ionising sources) smaller than unity and many ionising photons can escape freely along directions not covered by the gas \citep[e.g.][]{X18, X19, Schady2019}. (4) Finally, dust within the gas can also absorb part of the incident radiation. None of these effects are included in our model of the ionised gas; thus, the age estimate may be biased if we use the EW to compare models and observations.

We have performed the fitting with emission line ratios, which are not affected by the stellar continuum or the covering factor (the emission lines scale almost linearly with the covering factor). Yet, the line ratios still depend on the optical depth, dust within the gas, and abundance variations of individual elements. In this work, we assume the ionised gas is optically thick (i.e. radiation-bounded). This is reasonable for an active star-forming region where a huge amount of gas is required. Reducing the optical depth may affect the emission line ratios, since elements at different ionisation stages are located at different distances from the ionising sources.

The strong-line method \citep{o3n2.ref} has been widely used to estimate the gas-phase metallicity, which, however, has a large uncertainty of $\sim$0.18~dex. Since the nebular lines are optically thin, their fluxes are directly related to the amount of elements that produce them. Moreover, the collisionally excited lines are important coolants that can affect the thermal and ionisation structures. As a result, any abundance variations can affect the predicted line ratios. In practice, H~\textsc{ii} regions display a non-negligible variation in their abundance patterns \citep[e.g.][]{Pilyugin2003}. In this work, we find the H~\textsc{ii} region is nitrogen-enriched by a factor of 1.5 than the solar composition (Section~\ref{gas_model.sec}).

Dust within the gas may have complicated effects; the predicted line ratios vary with the grain types, grain size distribution and the dust-to-gas ratio \citep{Brinchmann2013}, all of which remain highly uncertain in different environments. As a simple test, we take the best-fitting (dust-free) model, add a dust component similar to that of the Orion Nebula, and then re-run \textsc{cloudy} to get its simulated spectrum. The line ratios change very little by $<$~0.1~dex in their logarithmic values (still within 1$\sigma$ uncertainties for [O~\textsc{iii}]/H$\beta$, [N~\textsc{ii}]/H$\alpha$ and [S~\textsc{ii}]/H$\alpha$ and within 1.5$\sigma$ uncertainties for [O~\textsc{i}]/H$\alpha$). This is possibly because the observed H~\textsc{ii} region has a very low ionisation parameter; \citet{Bottorff1998} find that dust has a large effect with high ionisation parameters but can absorb very little of the incident radiation at low ionisation levels. Thus, it is still valid to compare the observed line ratios with the dust-free models; assuming no dust greatly reduces the computing time.

One needs to correctly identify the relationship between the SN progenitor and stellar populations in the environment. It should be cautioned that an H~\textsc{ii} region may expand to a very large extent beyond the position of the ionising sources. In this work, for instance, the H~\textsc{ii} region has an extended envelope, which covers the position of SN~2004dg. However, this SN has a clear offset from the ionising stars in the star-forming complex. Thus, one cannot simply assume the SN progenitor is coeval with the ionising stars without carefully analysing their spatial distributions. Even if they are spatially coincident along the line of sight, one still needs to assess whether they are in chance alignment or physically associated (SN~2012P serves as a counter-example). This cannot be achieved without the help of high-spatial resolution imaging.

\subsubsection{High-spatial resolution imaging and the resolved/unresolved stellar populations}
\label{star_comments.sec}

High-spatial resolution imaging can reveal the brightest individual stars and compact star clusters in the environments of SNe out to several tens of Mpc. Star clusters are important sites of star formation \citep{L03.ref}; they are usually very bright and can be observed with high SNRs. As a well developed technique, cluster properties can be reliably derived by comparing its SED with model stellar populations.

Analysis of the resolved individual stars is much more complicated since they are often formed with a prolonged star formation history. This is further complicated by any dust between them along the line of sight such that the individual stars may have different extinctions. Fainter than star clusters, the individual stars may have larger photometric uncertainties and suffer from detection incompleteness. Thus, sophisticated techniques are required to derive their ages and extinctions.

In the near field, a large collection of resolved stars can be modelled based on binned CMDs \citep[e.g.][]{Harris2001, Rubele2018}. The general idea is that an observed CMD is the sum of partial CMDs arising from single-aged stellar populations (the weights of the partial CMDs provide insights into the star formation history). This method requires a significant number of detected stars since, otherwise, the colour-magnitude bins would contain too few stars and suffer from stochastic sampling effects. Due to the same reason, this method does not have a good accuracy for the age of young stellar populations (typically a few times 0.1~dex in log-age);  massive stars, which are used to distinguish them from the older ones, are much fewer than lower-mass stars.

Thus, this method is not very efficient in the research of SN environments. Even the ``nearby" SNe have distances of several or several tens of Mpc, where one can only detect a handful of the most massive stars at the high-mass end of the steep IMF \citep{imf.ref}. Moreover, since most core-collapse SNe have delay times of $<$~50~Myr (which could be longer if the pre-SN evolution involves binary interaction; \citealt{Z17}), a high age accuracy is needed for this age range to effectively constrain the progenitor properties and to distinguish different progenitor channels.

It is therefore very challenging to extract meaningful information for the complicated stellar populations from data with limited qualities. In this work, we use a Bayesian approach to derive ages of stellar populations by directly comparing the observed and predicted magnitudes of individual stars (Section~\ref{star_model.sec}). Suitable for SN environment studies, this method can derive accurate ages based on a limited number of stars. It can effectively use more than two filters and, perhaps more importantly, take advantage of any prior information from star clusters and IFU spectroscopy. Proper and reasonable priors can significantly improve the accuracy and resolve possible degeneracies  for the derived parameters.

\subsubsection{Combining IFU spectroscopy and high-spatial resolution imaging}

Thus, it is necessary to combine IFU spectroscopy and high-spatial resolution imaging to investigate SN environments. With the combined dataset, one can reveal both the gaseous and the stellar components in the SN environment and determine their metallicity, ages, extinctions, kinematics, projected spatial distribution, and even spatial distribution along the line of sight. Properties of the SN progenitor can be derived based on its host stellar population; with stellar evolutionary models, one can infer its pre-SN evolution and distinguish between different progenitor channels. A complete understanding of the SN environment may provide insight into any physical processes on a larger scale (e.g. spiral density waves) that control the star formation. Under different conditions, stars may be formed with different binary fractions/properties, which are directly related to the appearances of their SN explosions.

Analysis of the combined dataset can be carried out in an iterative Bayesian framework. One may start with reasonable assumptions and derive the parameters' posterior probabilities with less constraining priors. The validity of the assumptions should be re-assessed and the results for different components should be cross-checked. One can then update the assumptions, if necessary, and take advantage of any reasonable results from one component as new, more constraining priors for the another component. The posterior parameter distributions can be re-derived and this process can be iterated a few times until a fully self-consistent picture is reached.

\section{Summary and conclusions}

In this work, we present a detailed analysis of the environments of the Type~II-P SN~2004dg and Type~IIb SN~2012P based on spatially-resolved IFU spectroscopy by MUSE and high-spatial resolution imaging by HST. The two SNe occurred on one of the spiral arms of their host galaxy NGC~5806, where a populous star-forming complex is apparent with a size of $\sim$300~pc. SN~2004dg shows an offset of $\sim$200~pc from the complex while SN~2012P is spatially coincident with a compact star cluster within it. The star-forming complex has photo-ionised a giant H~\textsc{ii} region with a high H$\alpha$ luminosity.

We derive the ages and extinctions of stellar populations in the environment by modelling the ionised gas, the star cluster and the resolved stars. The star-forming complex contains at least two young populations; in addition, there are relatively older populations in the vicinity of both SNe. The different (mean) extinctions suggest that the star-forming complex lies in the foreground of the ionised gas and in the background of the older stars. Their relative spatial distributions along the line of sight could possibly arise from sequentially triggered star formation as the spiral density wave swept through their positions, but more observations are required to confirm this scenario.

For SN~2004dg, the age of its host stellar population corresponds to a progenitor with $M_{\rm ini}$~=~\iipmass~$M_\odot$, in agreement with the upper limit derived from its pre-explosion images. For SN~2012P, we suggest it is most likely to arise from the older stellar population instead of the star-forming complex; the derived progenitor mass, $M_{\rm ini}$~=~\iibmass~$M_\odot$, is well consistent with nebular spectroscopy.

As in the case of SN~2012P, star formation bursts on small scales may appear correlated if they are controlled by any physical processes on larger scales; thus, there could be a high probability of chance alignment between older and lower-mass SN progenitors with significantly younger stellar populations. This effect may partly explain the progenitor mass discrepancy between light-curve modelling and environment analysis for stripped-envelope SNe.

The analysis of SNe~2004dg/2012P demonstrates the necessity of combining spatially-resolved IFU spectroscopy and high-spatial resolution imaging for comprehensive investigations of both the gaseous and the stellar components in nearby SN environments. This is required for the accurate determination of their properties and for the correct identification of their relationships with each other and with the SN progenitors.

\section*{Acknowledgements}

We are very grateful to the anonymous referee for carefully reviewing our paper. We thank Selma de Mink,  Stephen Justham and Mathieu Renzo for a helpful discussion on the possible evolutionary scenarios of SN~2012P and Alison Sills for discussing the dynamical evolution of binaries in star clusters. This work is based on observations made with the NASA/ESA HST and observations collected at the European Southern Observatory under ESO programme 097.B-0165(A). N-CS, JRM and PAC acknowledge the funding support from the Science and Technology Facilities Council. EZ acknowledges support from the the Swiss National Science Foundation Professorship grant (project number PP00P2 176868) and from the Federal Commission for Scholarships for Foreign Students for the Swiss Government Excellence Scholarship (ESKAS No.\ 2019.0091).


\section*{Data availability}

Data used in this work are all publicly available from the ESO data archive (\url{http://archive.eso.org}) and the Mikulski Archive for Space Telescope (\url{https://archive.stsci.edu}).

\appendix

\section{Magnitude offsets}
\label{offset.sec}

\begin{figure}
\centering
\includegraphics[width=0.8\linewidth, angle=0]{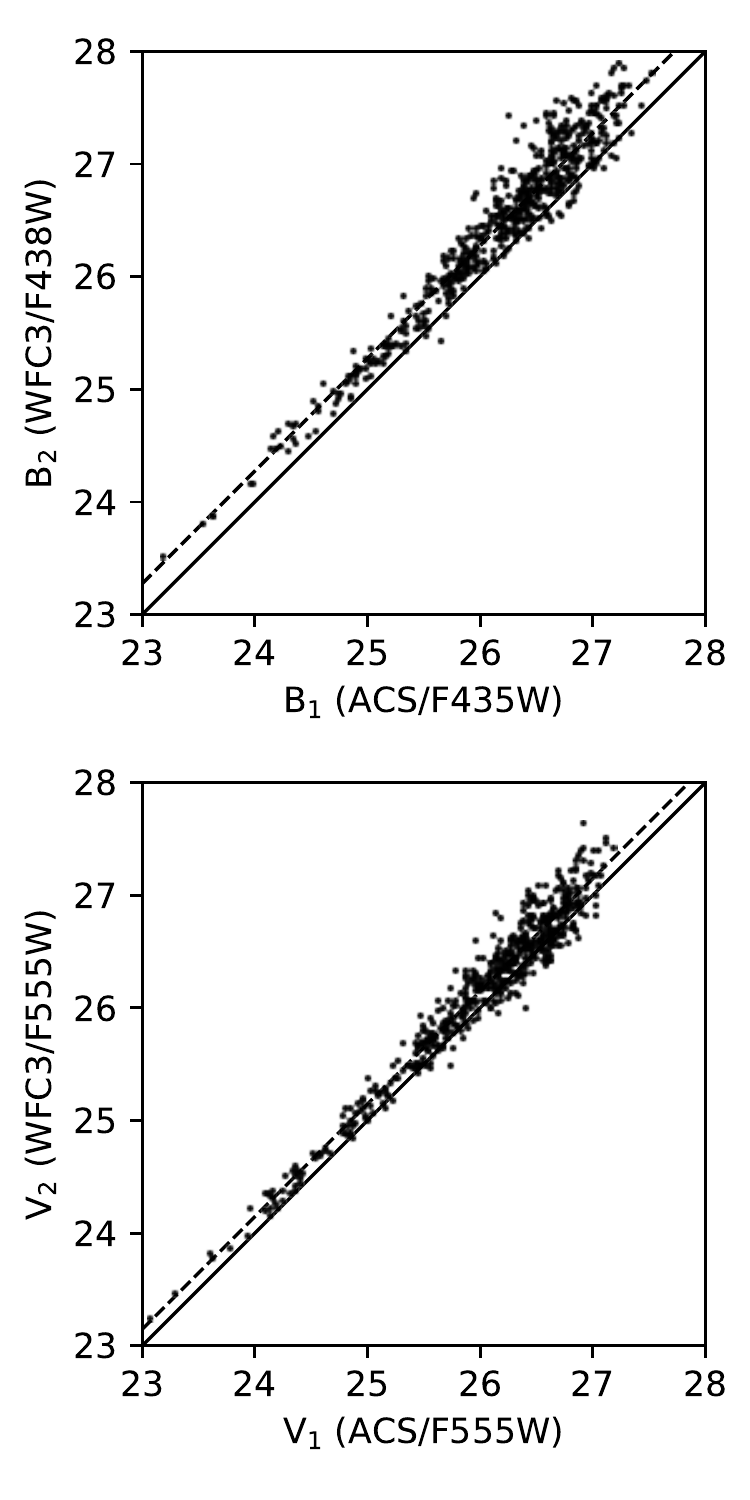}
\caption{Comparison between the B-band (top) and V-band (bottom) magnitudes converted from ACS and WFC3 observations.}
\label{abs1.fig}
\end{figure}

\begin{figure*}
\centering
\includegraphics[width=1\linewidth, angle=0]{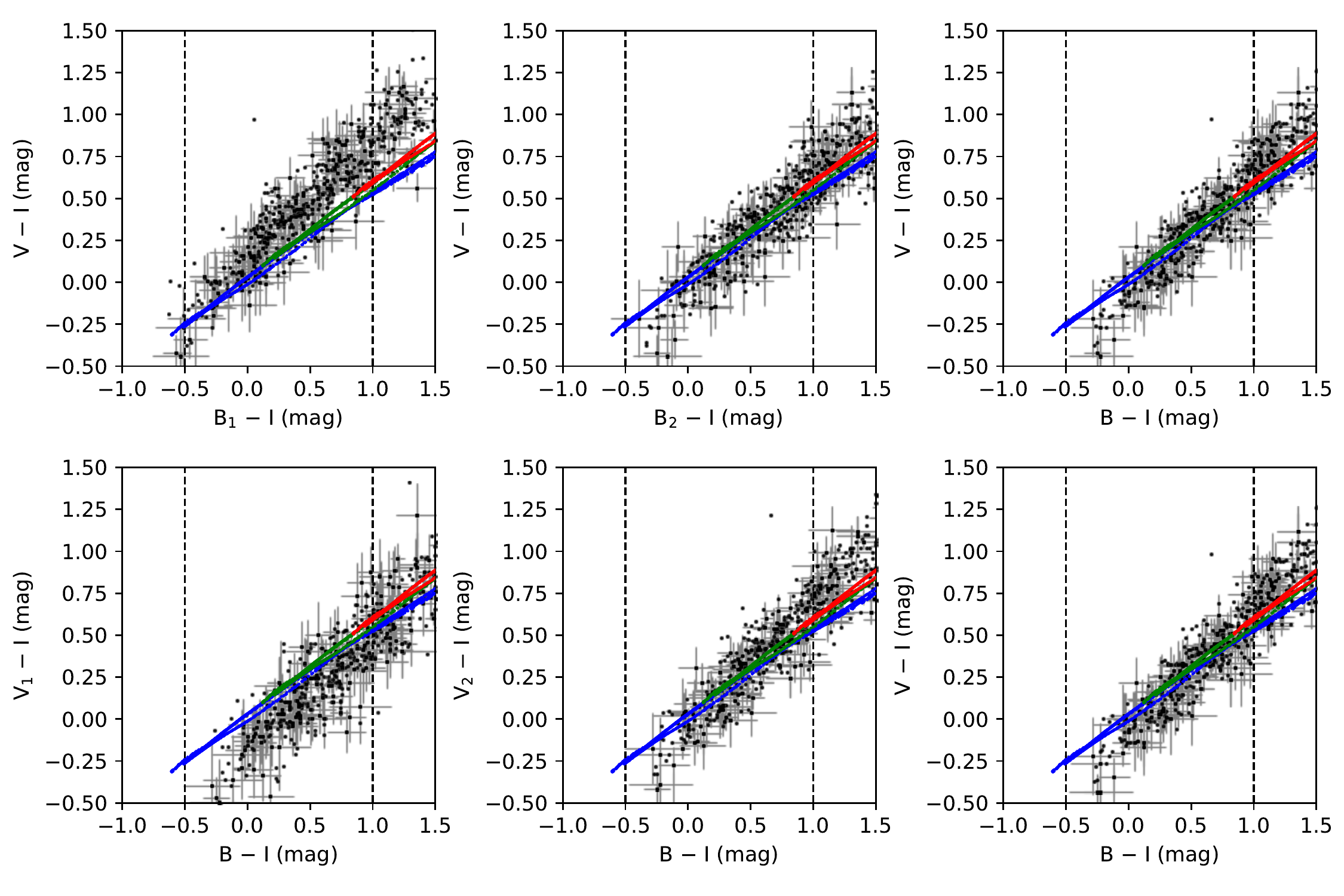}
\caption{Empirical correction of the B-band and V-band magnitudes in the colour-colour diagram (see text for details). The error bars reflect the photometric uncertainties; to reduce crowding, the error bars are displayed for only half of the data points (randomly choosen). The coloured curves are theoretical stellar locus from \textsc{parsec} isochrones with log($t$/yr)~=~6.8--8.0; they have been shifted by $A_V$=~0.0 (blue), 1.0 (green) and 2.0 (red). The vertical dashed lines enclose the colour range within which we compare the observed colours with the theoretical stellar locus to derive empirical corrections.}
\label{abs2.fig}
\end{figure*}

Small offsets are found when we compare the B- and V-band magnitudes converted from the ACS and WFC3 observations (Fig.~\ref{abs1.fig}). For simplicity, we shall use the subscript ``1" (``2") to denote magnitudes converted from ACS (WFC3) observations. B$_1$ and B$_2$ differ systematically by 0.274~mag while V$_1$ and V$_2$ differ by 0.146~mag. The magnitude offsets are independent of colour or magnitude. In contrast, ACS/F814W observations were also conducted at different epochs; however, no systematic offsets are discovered between different observations.

Similar magnitude offsets have also been found for HST data in a number of studies \citep[e.g.][]{Eldridge2015, S20a}. They might arise from imperfect photometry; \citet{Eldridge2015}, for instance, found that using different photometry parameters or different photometry packages can lead to different results. The offset in \citet{S20a} has an unknown origin, since they have carefully checked their photometry. In this work, it is not clear why there are offsets between the ACS and the WFC3 magnitudes. It is possible that, for instance, the model PSFs used may deviate from the true PSFs, which may vary with the telescope defocus at the time of observations.

It is not straightforward to see which set of photometry is closer to the ``correct" values; thus, we assume both the ACS and the WFC3 magnitudes have some systematic errors, which will then be derived empirically by comparing the observed source colours with theoretical stellar loci. We first take the average of V$_1$ and V$_2$ as the combined V-band magnitude (with inverse-variance weighting based on their photometric uncertainties). Since the offset between V$_1$ and V$_2$ is only 0.146~mag, their average should be very close to the ``correct" value with a difference of at most a few hundredth magnitude.

In the upper-left and upper-middle panels of Fig.~\ref{abs2.fig}, we show the sources on the (B$_1$~$-$~I, V~$-$~I) and (B$_2$~$-$~I, V~$-$~I) colour-colour diagrams. For comparison, we also display the theoretical stellar colours from \textsc{parsec} isochrones. The isochrones contain different ages [log($t$/yr)~=6.8--8.0] and interstellar extinctions ($A_V$~=~0.0, 1.0 and 2.0). In the range of $-$0.5~$<$~B~$-$~I~$<$~1.0~mag, however, the stellar locus has a very small dispersion. Thus, we can estimate the systematic errors in B$_1$ and B$_2$ by comparing the observed source colours with the theoretical stellar locus.

To do this, we fit a first-order polynomial to the stellar locus within $-$0.5~$<$~B~$-$~I~$<$~1.0~mag and, for each source within that range, derive its ``correct" B~$-$~I colour based on its V~$-$~I colour. We then calculate the average deviation between the observed and the derived colours with inverse-variance weighting. The average deviation is $-$0.282~mag for B$_1$ and $-$0.028~mag for B$_2$. The systematic errors are then subtracted from the observed magnitudes.

After this correction, B$_1$ and B$_2$ are combined together (B) with inverse variance weighting based on their photometric uncertainties. The upper-right panel of Fig.~\ref{abs2.fig} shows the colour-colour diagram with the combined magnitudes. It can be seen that the observed colours agree much better with the theoretical stellar locus.

With the updated B-band magnitude, we further check V$_1$ and V$_2$ on the colour-colour diagrams (lower-left and lower-middle panels of Fig.~\ref{abs2.fig}). Very similar to the above approach, we compare the observed colours with the theoretical stellar locus, now assuming the B-band magnitude is reliable and deriving corrections for V$_1$ and V$_2$. We find that V$_1$ is systematically brighter by 0.119~mag and V$_2$ is systematically fainter by 0.031~mag than the ``correct" value. V$_1$ and V$_2$ are then corrected and combined together. The lower-right panel of Fig.~\ref{abs2.fig} shows the sources' corrected colours, which are now very consistent with the theoretical ones.

The above steps can be iterated with updated magnitudes. This process converges very quickly. The final correction is $-$0.285~mag for B$_1$, $-$0.028~mag for B$_2$, $-$0.120~mag for V$_1$, and 0.030~mag for V$_2$ (which should be subtracted from the raw magnitudes). Thus, B$_1$ and V$_1$ have larger systematic errors while B$_2$ and V$_2$ are very close to the corrected magnitudes. After applying the corrections, the magnitudes from ACS and WFC3 can be combined together to increase the SNRs of the detected stars.

\section{Modelling of the resolved stars}
\label{fittingresolved.sec}

Modelling of the resolved stellar populations is based on \citet{Maund2016} with minor adaption. For a general single star $i$, its model magnitude in the $j$th band ($m^j_{\rm mod}$) can be predicted with its stellar parameters (stellar mass $M^i$, age $t^i$, and extinction $A_V^i$) based on \textsc{parsec} isochrones \citep{parsec.ref}. We calculate its likelihood function by comparing the model magnitudes with its observed photometric data ($\bm{D}^i$):
\begin{equation}
\label{cf1.eq}
p(\bm{D}^i \mid M^i, t^i, A_V^i) = \prod_j  p^i_j,
\end{equation}
where $p^i_j$ takes the form:
\begin{equation}
p^i_j = \dfrac{1}{\sqrt{2\pi} \, \sigma^j_{\rm obs}} \times\ {\rm exp} \left[ - \dfrac{1}{2} \left( \dfrac{m^j_{\rm mod} (M^i, t^i, A_V^i) - m^j_{\rm obs} }{\sigma^j_{\rm obs}} \right) ^2 \right],
\end{equation}
if the star is detected in the $j$th band with magnitude $m^j_{\rm obs}$ and uncertainty $\sigma^j_{\rm obs}$. In case the star is not detected, $p^i_j$ is substituted with
\begin{equation}
p^i_j = \dfrac{1}{2} \left[ 1 + {\rm erf} \left( \dfrac{m^j_{\rm mod} (M^i, t^i, A_V^i) -  m^j_{\rm lim}}{\sqrt{2} \, \sigma^j_{\rm lim}} \right) \right],
\end{equation}
where $m^j_{\rm lim}$ and $\sigma^j_{\rm lim}$ are the detection limit and its uncertainty for the $j$th band (estimated via artificial star tests; Section~\ref{phot.sec}). Since we are not interested in stellar mass, $M^i$ is treated as a nuisance parameter and can be marginalised over:
\begin{equation}
p(\bm{D}^i \mid t^i, A_V^i) = \int \left[ p(\bm{D}^i \mid M^i, t^i, A_V^i) \times p (M^i \mid t^i) \right] dM^i.
\end{equation}
We use the \citet{imf.ref} IMF as a prior, thus
\begin{equation}
p(M^i \mid t^i) \propto 
\begin{cases}
(M^i)^{-2.35} & \text{if $M^i_{\rm min}(t^i)$~$<$~$M^i$~$<$~$M^i_{\rm max}(t^i)$}, \\
0 & \text{otherwise};
\end{cases}
\end{equation}
where $M^i_{\rm min}(t^i)$ and $M^i_{\rm max}(t^i)$ are the minimum and maximum stellar mass of the isochrone for the age. The likelihood function should meet the normalisation requirement:
\begin{equation}
\int p(\bm{D}^i \mid t^i, A_V^i) \,dt^i \,dA_V^i = 1.
\end{equation}

The \textsc{parsec} isochrones are only for single stars. For a non-interacting binary, the total flux can be considered as the simple sum of fluxes of both stars. In this case, the model magnitudes can still be predicted, which depends on an additional parameter of the secondary-to-primary mass ratio ($q_i$). The likelihood function can be calculated similar to that for single stars; with a flat probability distribution between 0 and 1, $q_i$ is regarded as a nuisance parameter and can be marginalised over.

If we ignore interacting binaries and higher-order multiple systems, a detected source could be either a single star or a non-interacting binary. Assuming a binary fraction of $P_{\rm bin}$~=~0.5, its final likelihood function is the sum of these two scenarios:
\begin{multline}
\label{cf2.eq}
p(\bm{D}^i \mid t^i, A_V^i) = P_{\rm bin} \times p_{\rm bin}(\bm{D}^i \mid t^i, A_V^i) \\
+ (1 - P_{\rm bin}) \times p_{\rm sin}(\bm{D}^i \mid t^i, A_V^i).
\end{multline}

The observed stars can be considered as a mixture of model stellar populations, each characterised by an (mean) age and an (mean) extinction ($t^\prime_k$ and $A_V^{k \prime}$ for the $k$th population). In the Bayesian context, this forms a two-level hierarchical process: ($\underline{t^\prime}$, $\underline{A_V^\prime}$) are ``hyper-parameters" that govern the distribution of individual stars' parameters ($\underline{t}$, $\underline{A_V}$), which in turn are related to the photometric data $\underline{\bm{D}}$ via the likelihood function (the underlined quantities are vectors of parameters for all stars or stellar populations). Following the Bayes' theorem, the posterior probability distribution is 
\begin{multline}
p(\underline{t^\prime}, \underline{A_V^\prime}, \underline{t}, \underline{A_V} \mid \underline{\bm{D}}) \propto p(\underline{\bm{D}} \mid \underline{t}, \underline{A_V}) \\
\times p(\underline{t}, \underline{A_V} \mid \underline{t^\prime}, \underline{A_V^\prime}) \times p(\underline{t^\prime}, \underline{A_V^\prime});
\end{multline}
the three terms in the right side of the equation are the likelihood, the prior for individual stars' parameters, and the ``hyper-prior" for hyper-parameters of the model stellar populations. Since we are only interested in the model populations, the parameters for individual stars can be marginalised over
\begin{equation}
p(\underline{t^\prime}, \underline{A_V^\prime} \mid \underline{\bm{D}}) = \int p(\underline{t^\prime}, \underline{A_V^\prime}, \underline{t}, \underline{A_V} \mid \underline{\bm{D}}) \, d\underline{t} \, d\underline{A_V}.
\end{equation}
If we define $w_k$ as the probability of a detected source as a member of the $k$th population ($w_k$ can be treated as new hyper-parameters and fitted numerically), it can be shown that
\begin{multline}
p(\underline{t^\prime}, \underline{A_V^\prime} \mid \underline{\bm{D}}) = \left( \prod_i^{N_{\rm star}} \sum_k^{N_{\rm pop}} w_k p_{i, k} \right) \times p(\underline{t^\prime}, \underline{A_V}^\prime), \\
p_{i, k} = \int \left[ p(\bm{D}^i \mid t^i, A_V^i) \times p(t^i, A_V^i \mid t^{k \prime}, A_V^{k \prime}) \right] \,dt^i \, dA_V^i.
\end{multline}
Note that the likelihood $p(\bm{D}^i \mid t^i, A_V^i)$ can be derived with Equations~\ref{cf1.eq}--\ref{cf2.eq}. With pre-defined prior $p(t^i, A_V^i \mid t^{k \prime}, A_V^{k \prime})$ and hyper-prior $p(\underline{t^\prime}, \underline{A_V}^\prime)$, one can obtain the posterior probability distribution for the hyper-parameters. In this way, one can fit model stellar populations to the observed stars and estimate their ages and extinctions.


\bsp	
\label{lastpage}
\end{document}